\newif\ifapj
\apjtrue

\ifapj
  \documentclass[twocolumn]{aastex6}
\else
  \documentclass[12pt,preprint]{aastex}
\fi

\usepackage{amsmath,amsfonts,amssymb,enumerate,graphicx,
  longtable,float,times,textcomp,verbatim,url,nicefrac}
\def\apjl{ApJ}%
\def\apjs{ApJS}%
\def\ao{Appl.~Opt.}%
\def\aap{A\&A}%
\newcommand{\oiii}{[O{\sc iii}]\,$\lambda 5007$}
\newcommand{\heii}{He{\sc ii}\,$\lambda 4686$}
\newcommand{\angstrom}{\text{\normalfont\AA}}

\tabletypesize{\footnotesize}

\shorttitle{Reverberation Mapping in Five AGN}
\shortauthors{Fausnaugh et al.}

\begin{document}

\title{Reverberation Mapping of Optical Emission Lines in Five Active
  Galaxies}

\author{
  M.~M.~Fausnaugh\altaffilmark{1},
  C.~J.~Grier\altaffilmark{2,3},
  M.~C.~Bentz\altaffilmark{4},
  K.~D.~Denney\altaffilmark{1,5,6},
  G.~De Rosa\altaffilmark{7},
  B.~M.~Peterson\altaffilmark{1,5,7},
  C.~S.~Kochanek\altaffilmark{1,5},
  R.~W.~Pogge\altaffilmark{1,5},
  S.~M.~Adams\altaffilmark{8},
  A.~J.~Barth\altaffilmark{9},
  Thomas~G.~Beatty\altaffilmark{2,10},
  A.~Bhattacharjee\altaffilmark{11,12},
  G.~A.~Borman\altaffilmark{13},
  T.~A.~Boroson\altaffilmark{14},
  M.~C.~Bottorff\altaffilmark{15},
  Jacob~E.~Brown\altaffilmark{16},
  Jonathan~S.~Brown\altaffilmark{1},
  M.~S.~Brotherton\altaffilmark{11},
  C.~T.~Coker\altaffilmark{1},
  S.~M.~Crawford\altaffilmark{17},
  K.V.~Croxall\altaffilmark{18},
  Sarah~Eftekharzadeh\altaffilmark{11},
  Michael~Eracleous\altaffilmark{2,3,19},
  M.~D.~Joner\altaffilmark{20},
  C.~B.~Henderson\altaffilmark{21,22},
  T.~W.-S.~Holoien\altaffilmark{1,5},
  Keith~Horne\altaffilmark{23},
  T.~Hutchison\altaffilmark{15},
  Shai~Kaspi\altaffilmark{24},
  S.~Kim\altaffilmark{1},
  Anthea~L.~King\altaffilmark{25},
  Miao~Li\altaffilmark{26},
  Cassandra~Lochhaas\altaffilmark{1},
  Zhiyuan~Ma\altaffilmark{16},
  F.~MacInnis\altaffilmark{15},
  E.~R.~Manne-Nicholas\altaffilmark{4},
  M.~Mason\altaffilmark{11},
  Carmen~Montuori\altaffilmark{27},
  Ana~Mosquera\altaffilmark{28},
  Dale~Mudd\altaffilmark{1},
  R.~Musso\altaffilmark{15},
  S.~V.~Nazarov\altaffilmark{13},
  M.~L.~Nguyen\altaffilmark{11},
  D.~N.~Okhmat\altaffilmark{13},
  Christopher A. Onken\altaffilmark{29},
  B.~Ou-Yang\altaffilmark{4},
  A.~Pancoast\altaffilmark{30,31},
  L.~Pei\altaffilmark{9,32},
  Matthew~T.~Penny\altaffilmark{1,33},
  Rados\l{}aw~Poleski\altaffilmark{1},
  Stephen~Rafter\altaffilmark{34},
  E.~Romero-Colmenero\altaffilmark{17,35},
  Jessie~Runnoe\altaffilmark{2,3,36},
  David~J.~Sand\altaffilmark{37},
  Jaderson~S.~Schimoia\altaffilmark{38},
  S.~G.~Sergeev\altaffilmark{13},
  B.~J.~Shappee\altaffilmark{39,40,41},
  Gregory~V.~Simonian\altaffilmark{1},
  Garrett~Somers\altaffilmark{42,43},
  M.~Spencer\altaffilmark{20},
  D.~A.~Starkey\altaffilmark{23},
  Daniel~J.~Stevens\altaffilmark{1},
  Jamie~Tayar\altaffilmark{1},
  T.~Treu\altaffilmark{44},
  Stefano~Valenti\altaffilmark{45}, 
  J.~Van Saders\altaffilmark{39,40},
  S.~Villanueva Jr.\altaffilmark{1},
  C.~Villforth\altaffilmark{46},
  Yaniv~Weiss\altaffilmark{47},
  H.~Winkler\altaffilmark{48},
  W.~Zhu\altaffilmark{1}
}

\altaffiltext{1}{Department of Astronomy, The Ohio State University,
  140 W 18th Ave, Columbus, OH 43210, USA}
\altaffiltext{2}{Department
  of Astronomy \& Astrophysics, The Pennsylvania State University, 525
  Davey Laboratory, University Park, PA 16802, USA}
\altaffiltext{3}{Institute for Gravitation and the Cosmos, The Pennsylvania State University, University Park, PA 16802, USA}
\altaffiltext{4}{Department of Physics and Astronomy, Georgia State University, Atlanta,
GA 30303}
\altaffiltext{5}{Center for Cosmology and AstroParticle Physics, The
  Ohio State University, 191 West Woodruff Ave, Columbus, OH 43210,
  USA}
\footnotetext[6]{NSF Postdoctoral Research Fellow}
\altaffiltext{7}{Space Telescope Science Institute, 3700 San Martin Drive, Baltimore, MD 21218, USA}
\altaffiltext{8}{Cahill Center for Astrophysics, California Institute of Technology, Pasadena, CA 91125, USA}

\altaffiltext{9}{Department of Physics and Astronomy, 4129 Frederick
Reines Hall, University of California, Irvine, CA 92697, USA}

\altaffiltext{10}{Center for Exoplanets and Habitable Worlds, The Pennsylvania State University, 525 Davey Lab, University Park, PA 16802}

\altaffiltext{11}{Department of Physics and Astronomy, University of
Wyoming, 1000 E. University Ave. Laramie, WY, USA}

\altaffiltext{12}{The Department of Biology, Geology, and Physical Sciences, Sul Ross State
University, WSB 216, Box-64, Alpine, TX, 79832, USA}

\altaffiltext{13}{Crimean Astrophysical Observatory, P/O Nauchny,
Crimea 298409, Russia}

\altaffiltext{14}{Las Cumbres Global Telescope Network, 6740 Cortona Drive, Suite 102,
Santa Barbara, CA  93117, USA}

\altaffiltext{15}{Fountainwood Observatory, Department of Physics FJS 149,
Southwestern University, 1011 E. University Ave., Georgetown, TX 78626, USA}

\altaffiltext{16}{Department of Physics and Astronomy, University of Missouri, Columbia, USA}

\altaffiltext{17}{South African Astronomical Observatory, P.O. Box
  9, Observatory 7935, Cape Town, South Africa}

\altaffiltext{18}{Illumination Works LLC., 5550 Blazer Parkway, Suite
  152 Dublin, OH 43017}

\altaffiltext{19}{Center for Relativistic Astrophysics, Georgia Institute of Technology, Atlanta, GA 30332, USA}

\altaffiltext{20}{Department of Physics and Astronomy, N283 ESC, Brigham Young University,
Provo, UT 84602-4360, USA}

\altaffiltext{21}{Jet Propulsion Laboratory, California Institute of Technology, 4800 Oak Grove Drive, Pasadena, CA 91109, USA}
\footnotetext[22]{NASA Postdoctoral Program Fellow}

\altaffiltext{23}{SUPA Physics and Astronomy, University of
St. Andrews, Fife, KY16 9SS Scotland, UK}

\altaffiltext{24}{School of Physics and Astronomy, Raymond and Beverly Sackler Faculty of Exact
Sciences, Tel Aviv University, Tel Aviv 69978, Israel}

\altaffiltext{25}{School of Physics, University of Melbourne, Parkville, VIC 3010, Australia}

\altaffiltext{26}{Department of Astronomy, Columbia University, 550 W120th Street, New York, NY 10027, USA}
\altaffiltext{27}{DiSAT, Universita dell'Insubria, via Valleggio 11, 22100, Como, Italy}

\altaffiltext{28}{Physics Department, United States Naval Academy, Annapolis, MD 21403}

\altaffiltext{29}{ Research School of Astronomy and Astrophysics, Australian National University, Canberra, ACT 2611, Australia}

\altaffiltext{30}{Harvard-Smithsonian Center for Astrophysics, 60 Garden Street, Cambridge, MA 02138, USA}
\footnotetext[31]{Einstein Fellow}

\altaffiltext{32}{Department of Astronomy, University of Illinois at Urbana-
Champaign, Urbana, IL 61801, USA}

\footnotetext[33]{Sagan Fellow}
\altaffiltext{34}{Department of Physics, Faculty of Natural Sciences, University of Haifa,
   Haifa 31905, Israel}

\altaffiltext{35}{Southern African Large Telescope Foundation,
  P.O. Box 9, Observatory 7935, Cape Town, South Africa}

\altaffiltext{36}{Department of Astronomy, University of Michigan,
  1085 S. University, 311 West Hall, Ann Arbor, MI 48109-1107}

\altaffiltext{37}{Physics \& Astronomy Department, Texas Tech University, Box 41051, Lubbock, TX 79409-1051, USA}

\altaffiltext{38}{Instituto de F\'isica, Universidade Federal do Rio Grande do Sul, Campus do Vale, Porto Alegre, RS, Brazil}

\altaffiltext{39}{Carnegie Observatories, 813 Santa Barbara Street, Pasadena, CA 91101, USA}
\footnotetext[40]{Carnegie-Princeton Fellow}
\footnotetext[41]{Hubble Fellow}
\altaffiltext{42}{Department of Physics and Astronomy, Vanderbilt University, 6301 Stevenson Circle, Nashville, TN, 37235}
\altaffiltext{43}{VIDA Postdoctoral Fellow}

\altaffiltext{44}{Department of Physics and Astronomy, University of
California, Los Angeles, CA 90095-1547, USA}

\altaffiltext{45}{Department of Physics, University of California, One
  Shields Avenue, Davis, CA 95616, USA}

\altaffiltext{46}{University of Bath, Department of Physics, Claverton Down, BA2 7AY, Bath, United Kingdom}

\altaffiltext{47}{Physics Department, Technion, Haifa 32000, Israel}

\altaffiltext{48}{Department of Physics, University of Johannesburg,
  PO Box 524, 2006 Auckland Park, South Africa}

\begin{abstract}
  We present the first results from an optical reverberation mapping
  campaign executed in 2014, targeting the active galactic nuclei
  (AGN) MCG+08-11-011, NGC\,2617, NGC\,4051, 3C\,382, and
  Mrk\,374. Our targets have diverse and interesting observational
  properties, including a "changing look" AGN and a broad-line radio
  galaxy.  Based on continuum-H$\beta$ lags, we measure black hole
  masses for all five targets.  We also obtain H$\gamma$ and \heii\
  lags for all objects except 3C\,382.  The \heii\ lags indicate
  radial stratification of the BLR, and the masses derived from
  different emission lines are in general agreement.  The relative
  responsivities of these lines are also in qualitative agreement with
  photoionization models.  These spectra have extremely high
  signal-to-noise ratios (100--300 per pixel) and there are excellent
  prospects for obtaining velocity-resolved reverberation signatures.
\end{abstract}

\section{Introduction}
Understanding the interior structure of active galactic nuclei (AGN)
has been a major goal of extragalactic astrophysics since their
identification as cosmological objects \citep{Schmidt1963}.  The
current schematic structure of the central part of an AGN includes
three main components: an accretion disk around a super-massive black
hole (SMBH), a broad line region (BLR), and an obscuring structure at
some distance beyond the BLR.
This basic picture accounts for the large luminosities and prominent
recombination/excitation lines observed in Seyfert galaxy and quasar
spectra \citep{Burbidge1967, Weedman1977}, as well as the dichotomy
between Type 1 and Type 2 objects \citep{Lawrence1991,Antonucci1993}.

While this model has qualitatively explained the observational
properties of AGN, the details of AGN interior structure remain poorly
understood.  The basic physics of the accretion disk are probably
linked to the magnetorotational instability \citep{Balbus1998}, but it
has not been possible to fully simulate an accretion disk and compare
with observations \citep{Koratkar1999, Yuan2014}.  It is also unclear
if the BLR simply consists of ambient gas near the SMBH, or if it is
more directly connected with the accretion process.  For example,
broad-line emitting gas might correspond to inflowing gas from large
scales that feeds the accretion disk,
or a portion of the BLR gas may be the result of an outflowing wind
driven by radiation pressure from the accretion disk
\citep{Collin-Souffrin1987, Murray1997, Elvis2000, Proga2004,
  Proga2010, Higginbottom2014, Elitzur2016}. The BLR could instead
correspond to the portion of the obscuring structure lying within the
dust sublimation radius \citep{Netzer1993, Simpson2005, Gaskell2008,
  Nenkova2008, Mor2012}.  Other models explore the possibility that
the accretion disk, BLR, and obscuring structure are not distinct at
all, but different observational aspects of a single structure bound
to the central SMBH (e.g., \citealt*{Elitzur2006, Czerny2011,
  Goad2012}).

Reverberation mapping (RM, \citealt{Blandford1982, Peterson1993,
  Peterson2014}) is an effective way of investigating these scenarios.
RM exploits the intrinsic variability of AGN to investigate the matter
distribution around the SMBH.  The inner parts of the accretion disk
emit in the far/extreme UV, providing ionizing photons that drive line
emission from BLR gas. As the accretion disk stochastically varies,
changes in the continuum flux are reprocessed as line emission by BLR
gas after a time delay that corresponds to the light-travel time
across the BLR.  Measuring this time delay (or ``lag'') provides a
means of measuring the characteristic size-scale of the line-emitting
gas.  Similarly, the UV continuum (or X-rays) deposits a small
fraction of the accretion luminosity in the outer parts of the
accretion disk and obscuring structure. Continuum variations will
therefore change the local temperature of these structures, which can
drive variable emission at longer continuum wavelengths---the outer
part of the accretion disk emits primarily in the optical and the
obscuring structure emits in the IR.  By measuring any lag between the
primary UV signal and light echoes at longer wavelengths, it is
possible to ``map'' the size of the accretion disk and obscuring
structure.

Early RM experiments were able to measure or constrain the physical
scales of the three primary components: the accretion disk is of order
a few light days from the SMBH (e.g.,
\citealt{Wanders1997,Sergeev2005}), the BLR ranges from several light
days to a few light months or light years, depending on the AGN
luminosity (\citealt{Wandel1999, Kaspi2000, Peterson2004, Kaspi2005}),
and the obscuring structure extends several light months or light
years beyond the BLR \citep{Clavel1989, Oknyanskij2001,Suganuma2006}.
More recent RM studies have provided additional details.  The
detection of continuum lags across the accretion disk provides
information about the disk's temperature gradient, and it appears that
the disks are somewhat larger than the predictions from standard
models (e.g., \citealt{Shappee2014, Edelson2015, Fausnaugh2016,
  McHardy2016}), as also found in microlensing studies of lensed
quasars (e.g., \citealt{Morgan2010,Blackburne2011,Mosquera2013}).
Mid- to far-IR echoes from the obscuring structure have facilitated
investigation of AGN dust properties, and suggest that the obscuring
structure is clumpy and has a mixed chemical composition
\citep{Kishimoto2007, Vazquez2015}.

RM of the BLR is of particular importance for AGN studies because
velocity information in the broad-line profile combined with the
observed time delay provides a well-calibrated estimate of the SMBH
mass.  Approximately 60 AGN have RM mass measurements
\citep{Bentz2015}, and this sample anchors the scaling relations used
to infer the majority of SMBH masses throughout the universe (e.g.,
\citealt{McLure2004, Vestergaard2006, Trakhtenbrot2012, Park2013,
  Mejia-Restrepo2016}, and references therein).  New insights into the
BLR structure have also become available with velocity-resolved
analyses (e.g., \citealt{Denney2010, Bentz2010, Barth2015,
  Valenti2015,Du2016}).  By combining information about the BLR time
delay as a function of line-of-sight velocity, it is possible to
distinguish among geometric and dynamical configurations, such as
flattened versus spherical matter distributions and dynamics dominated
by rotation, infall, or outflow \citep{Horne1994, Horne2004,
  Bentz2010, Grier2013, Pancoast2014a, Pancoast2014b}.  So far, only
about 10 AGN have such detailed velocity-resolved results, but they
suggest a wide range of dynamics and geometries.

In this work, we present the first results from an intensive RM
campaign executed in 2014.  This campaign had two primary goals: to
measure SMBH masses in several objects with interesting or peculiar
observational properties, and to expand the sample of AGN with
velocity-resolved reverberation signatures.  NGC\,5548 was also
observed in this campaign as part of the multiwavelength AGN STORM
project (\citealt{DeRosa2015, Edelson2015, Fausnaugh2016, Goad2016}).
Ground-based spectroscopic results for this object are presented by
\citet{Pei2017}.  Here, we present the final data and initial analysis
of other AGN from this campaign, reporting continuum and line light
curves, continuum-line lag measurements, and SMBH masses for five
objects.  We detected variability in the H$\beta$, H$\gamma$ and
\heii\ emission lines for most objects, which we also use to explore
the photoionization conditions in the BLR
\citep{Korista2004,Bentz2010}.  These data are of exceptional quality
and should allow us recover velocity-resolved reverberation signatures
in future work.

In \S2, we present our target AGN, observations, data reduction, and
light curves.  In \S3, we explain our time-series analysis and report
continuum-line lags.  In \S4 we measure the gas velocities and
estimate SMBH masses.  In \S5 we discuss our results, and in \S6 we
summarize our findings.  We assume a consensus cosmology with $H_0 =
70 {\rm\ km\ s^{-1}\ Mpc^{-1}}$, $\Omega_{\rm m} = 0.3$, and
$\Omega_{\Lambda} = 0.7$.

\section{Observations and Data Reduction}
\subsection{Targets}
\floattable
\begin{deluxetable}{lrrcrcccc}
\tablewidth{0pt}
\tablecaption{Target Properties \label{tab:targets}}
\tablehead{
\colhead{Object} & \colhead{Redshift}& \colhead{$D_L$} &  \colhead{Number of} &\colhead{$F$ [O{\sc iii}]$\lambda$5007 }  & \colhead{[O{\sc iii}]$\lambda$5007 } & \colhead{$\log \lambda L_{5100\angstrom}$} & \colhead{$\log \lambda L_{\rm host}$} & \colhead{{\it E}({\it B}$-${\it V})}\\
 & &\colhead{(Mpc)} & \colhead{Good-weather Epochs} & \colhead{($10^{-14}$ erg cm$^{-2}$ s$^{-1}$)} & \colhead{Light Curve Scatter (\%)} & \colhead{[erg s$^{-1}$]} &\colhead{[erg s$^{-1}$]} & \colhead{(mag)}\\
\colhead{(1)}&\colhead{(2)}&\colhead{(3)}&\colhead{(4)}&\colhead{(5)}&\colhead{(6)}&\colhead{(7)}&\colhead{(8)}&\colhead{(9)}
}
\startdata
MCG+08-11-011 & 0.0205 &89.1 & 6 & $ 61.33 \pm  0.21$ & 0.09 & 43.59 & 43.28 & 0.19 \\
NGC 2617      & 0.0142 &61.5 & 3 & $ 6.88 \pm  0.06$ & 1.37 & 43.12 & 42.95 & 0.03 \\
NGC 4051      & 0.0023 &17.1 & 3 & $ 41.30 \pm  0.26$ & 0.36 & 42.38 & 42.23 & 0.01 \\
3C 382        & 0.0579 &258.7 & 6 & $ 7.87 \pm  0.06$ & 0.92 & 44.20 & 43.98 & 0.06 \\
Mrk 374       & 0.0426 &188.5 & 3 & $ 7.27 \pm  0.11$ & 0.62 & 43.98 & 43.61 & 0.05 \\
\enddata
\tablecomments{Column 2 is taken from the NASA Extragalactic Database.
  Column 3 gives the luminosity distance $D_L$ in a concensus
  cosmology, except for NGC\,4051 for which the luminosity distance is
  from \citet{Tully2008}.  Column 4 gives the number of nights with
  clear and stable conditions on which each object was observed.  Each
  object had three observations per night, which were used to
  calculate the narrow [O{\sc iii}]$\lambda$5007 line flux.  The line
  flux and its uncertainty are given in Column 5.  Column 6 gives the
  fractional variation of the [O{\sc iii}]$\lambda$5007 line light
  curve, which serves as an estimate of the night-to-night calibration
  error (\S2.5.1).  Column 7 gives the observed luminosity (corrected
  for Galactic extinction), calculated from the observed 5100\,\AA\
  rest-frame light curve and Column 3.  Column 8 gives the luminosity
  of the host-galaxy starlight in the spectroscopic extraction
  aperture, also corrected for Galactic extinction (\S5.1).  Note that
  Column 7 includes the contribution from the host galaxy.  Column 9
  gives the Galactic reddening value from \citet{Schlafly2011}.}
\end{deluxetable}

In spring of 2014 we monitored 11 AGN over the course of a six-month
RM campaign.  The AGN were selected with the aim of expanding the
database of RM SMBH masses, particularly for objects with diverse and
peculiar observational characteristics.  The second goal of our
campaign was to investigate the dynamics and geometry of the BLR with
velocity-resolved reverberation signatures, i.e., velocity-delay maps
and dynamical models (see e.g. \citealt{Grier2013, Pancoast2014a}).
Here, we focus on results related to SMBH masses, and we will pursue
the velocity-resolved analysis in future work.

\begin{figure*}
\includegraphics[width=\textwidth]{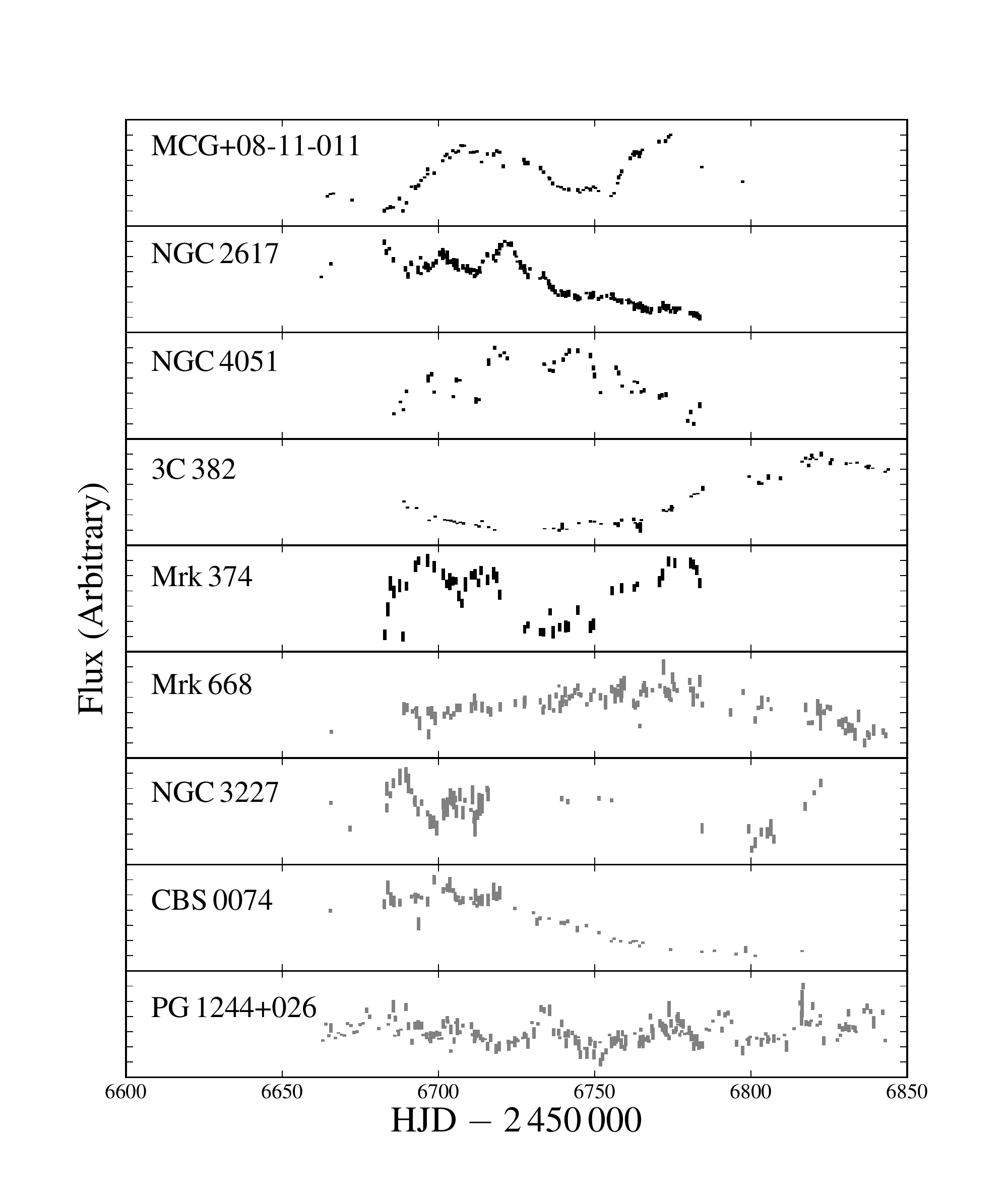}
\caption{{\it g}-band light curves of all targets in the 2014
  monitoring campaign except for Akn 120 and NGC 5548 (see \S2.1).  In
  this study, we focus on MCG+08-11-011, NGC\,2617, NGC\,4051,
  3C\,382, and Mrk\,374.  The extent of the errorbars on the open
  white circles represent 10\% variations for each flux
  scale.\label{fig:targets}}
\end{figure*}

Figure\,\ref{fig:targets} shows {\it g}-band light curves from the Las
Cumbres Observatory (LCO) 1m network for nine of our targets (we
discuss these data in detail in \S2.3).  Not shown are Akn\,120, which
was dropped early in the campaign because of low variability, and
NGC\,5548, for which the results are presented elsewhere
(\citealt{Fausnaugh2016,Pei2017}).  In order to estimate a black hole
mass, we must measure a continuum--line lag.  We have not been able to
measure such a reverberation signal for Mrk\,668, NGC\,3227,
CBS\,0074, and PG\,1244+026.  These sources have lower
  signal-to-noise ratios (S/Ns) than the other objects (generally
  30--70 per pixel, although NGC\,3227 was $\sim\! 90$ per pixel; see
  \S2.5.3), and they display lower variability amplitudes.  The
  fractional root-mean-square amplitude ($F_{\rm var}$ as defined in
  \S2.5.3 below) is 0.012 for Mrk\,668, 0.037 for NGC\,3227, 0.010 for
  CBS\,0074, and 0.025 for PG\,1244+026.  For Mrk\,668, the slow rate
  of change in the light curve also makes it impossible to measure
  short lags.  For NGC\,3227, the light curve is problematic because
of the limited sampling and large gaps; however, this object was also
observed during a monitoring campaign in 2012, and we will combine the
data from both campaigns in a future analysis.  For CBS\,0074 and
PG\,1244+026, we have not been able to obtain a sufficiently precise
calibration of the spectra (see \S2.2.2) to detect emission line
variability.

We succeeded in measuring black hole masses for MCG+08-11-011,
NGC\,2617, NGC\,4051, 3C\,382, and Mrk\,374.  Table\,\ref{tab:targets}
lists the some of the important properties of these objects (several
of which are measured in this study), and we provide additional
comments as follows:

\begin{enumerate}[i.]
\item MCG+08-11-011 is a strong X-ray source for which spectral
  signatures of a relativistically-broadened Fe K$\alpha$ line have
  been observed with {\it Suzaku} \citep{Bianchi2010}.  The Fe
  K$\alpha$ emission is believed to be emitted close to the inner edge
  of the accretion disk, and can potentially be used to measure the
  spin parameter of the black hole.  Because the black hole mass and
  spin are to some extent degenerate when fitting the broad Fe
  K$\alpha$ profile, an independent mass estimate from RM can greatly
  assist with the spin measurement.

\item NGC\,2617 was discovered by \citet{Shappee2014} to be a
  ``changing look'' AGN.  In 2013, after a large X-ray/optical
  outburst, follow-up spectroscopic observations showed the presence
  of broad lines, while archival spectra from 2003 show only a weak
  broad component of H$\alpha$.  This means that the classification of
  NGC\,2617 changed from a Seyfert 1.9 to Seyfert 1.0 sometime in the
  intervening decade.  Few optical ``changing look'' AGN are known,
  although systematic searches through long-term survey data (such as
  the SDSS, \citealt{LaMassa2015,MacLeod2016}) and targeted repeat
  spectroscopy \citep{Runnoe2016,Runco2016, Ruan2016} have recently
  expanded the sample size to approximately 20 objects, depending on
  how ``changing look'' AGN are defined.  The absolute rate of this
  phenomenon is very uncertain, but these recent studies suggest that
  it may be relatively common over several decades, a time scale that
  long-term spectroscopic surveys are only beginning to probe.
  Velocity-resolved dynamical information is of special interest in an
  object such as this, since the presence of outflows or infall may
  provide clues about the physical mechanism behind the change in
  Seyfert category.

\item NGC\,4051 has been the target of several optical and X-ray RM
  campaigns \citep{Shemmer2003,Peterson2000, Peterson2004,
    Denney2009b,Miller2010,Turner2017}.  However, the short H$\beta$
  lag, comparable to the cadence of most monitoring campaigns, has led
  to mixed and inconsistent results.  \citet{Denney2009b} found an
  H$\beta$ lag of $1.87 \pm 0.52$ days, roughly a factor of 2 smaller
  than previous studies.  Because of the large change, as well as the
  lag's small value compared to the monitoring cadence, we re-observed
  NGC\,4051 during the 2014 campaign to check this result.  For one
  month of the campaign (2014 February 17 to 2014 March 16 UTC), we
  also increased the monitoring cadence of NGC\,4051 to twice nightly,
  in order to securely resolve the expected short H$\beta$ lag.

  NGC\,4051 is also an archetypal narrow-line Seyfert 1 (NLS1),
  meaning that the width of its H$\beta$ line is $\lesssim 2\,000
  {\rm\ km\ s}^{-1}$.  There are two competing theories to explain the
  NSL1 phenomenon: high accretion rates or rotationally-dominated BLR
  dynamics seen nearly face-on.  Both explanations can account for the
  narrow linewidths given the AGN luminosity.  Insight into the
  structure of the BLR can help distinguish between these
  explanations, so there is considerable interest in reconstructing a
  velocity-delay map for this object.

\item 3C\,382 is an FR II broad-line radio galaxy
  \citep{Osterbrock1975,Osterbrock1976}.  Few radio-loud AGN have RM
  mass measurements, although there are notable examples such as
  3C\,390 \citep{Shapovalova2010, Dietrich2012}, 3C\,273
  \citep{Kaspi2000, Peterson2004}, and 3C\,120 \citep{Peterson2004,
    Grier2012}.  These objects are typically more luminous than
  radio-quiet AGN, so they have large lags (of order months to years)
  that are difficult and expensive to measure.  However, radio
  emission is thought to be associated with more massive black holes,
  which can be tested by anchoring radio-loud AGN to the RM mass
  scale.  Radio jets can also provide an indirect estimate of the
  inclination of the BLR, if the BLR is a disky structure with the
  rotation axis aligned to that of the jet (\citealt{Wills1986}).
  Several jet-orientation indicators exist for 3C\,382, and
  \citet{Eracleous1995} estimated the BLR inclination in 3C\,382 using
  dynamical models of the double-peaked H$\alpha$ profile.
  Velocity-delay maps and dynamical models would provide an
  interesting comparison to these estimates.

\item We observed Mrk\,374 in an RM campaign from 2012, but the AGN
  did not display sufficient variability to measure emission line lags
  at that time.  Although Mrk\,374 is our least variable source, we
  succeeded in measuring a line lag from the 2014 campaign, and we
  present the first RM-based black hole mass here.
\end{enumerate}

\subsection{Spectra}

\subsubsection{Observations}
We obtained spectra on an approximately daily cadence between 2014
January 04 and 2014 July 06 UTC using the Boller and Chivens CCD
Spectrograph on the 1.3m McGraw-Hill telescope at the MDM Observatory.
We used the 350 mm$^{-1}$ grating, yielding a dispersion of 1.33\,\AA\
per pixel with wavelength coverage from 4300\,\AA\ to 5600\,\AA.  We
kept the position angle of the slit fixed to 0$^\circ$ for the entire
campaign, with a slit width of 5\farcs0 to minimize losses due to
differential refraction and aperture effects caused by extended
emission (i.e., the host-galaxy and narrow line region,
\citealt{Peterson1995}).  Because of the large slit width, the
spectroscopic resolution for point sources (such as the AGN) is
limited by the image seeing.  We discuss this in more detail in \S4,
but comparison with high-resolution observations suggest that the
effective spectral resolution is approximately 7.0\,\AA.

The two-dimensional spectra were reduced using standard {\tt IRAF}
tasks for overscan, bias, and flat-field corrections, and cosmic rays
were removed using LA-cosmic \citep{vanDokkum2001}.  We extracted
one-dimensional spectra from a 15\farcs0 window centered on a linear
fit to the trace, and we derived wavelength solutions from comparison
lamps taken in the evening and morning of all observing nights.  We
also corrected for zero-point shifts in the wavelength solutions (due
to flexure in the telescope) by taking xenon lamp exposures just prior
to each observing sequence.  However, every AGN was observed for a
series of three 20 minute exposures and the wavelength zero-point can
drift over the course of this hour, especially at high airmass.  We
therefore tie the wavelength solution of the first exposure to the
contemporaneous xenon lamp, and then apply shifts that align the
\oiii\ emission line of subsequent exposures to that of the first.
This procedure results in wavelength solutions accurate to 0.56\,\AA,
as measured from night-sky emission lines.

We applied relative flux calibrations using sensitivity curves derived
from nightly observations of standard stars.  For most of the
campaign, we use Feige 34 \citep{Oke1990} to define the nightly
sensitivity curve.  However, this star began to set near dusk at the
end of the campaign, so we tied our relative flux calibration to
BD+33$^{\circ}$2642 \citep{Oke1990} for the final two weeks. The
change in standard star could potentially result in a systematic
change in the observed continuum slopes.  However, BD+33$^{\circ}$2642
and Feige 34 were observed for a one-month overlap period before the
transition, and the sensitivity curves derived from both stars agree
well during this time period.  Of the targets presented here, only
3C\,382 was observed during the final two weeks, and we did not find
any anomalous changes in the spectral slope during this period.
As a check on the relative flux calibration, we also looked
  for a ``bluer when brighter'' trend, caused by an increasing
  fraction of host-galaxy light when the AGN is in a faint state
  and/or intrinsic variations in the AGN spectral energy distribution
  (e.g., \citealt{Wilhite2005, Sakata2010}).  We measured the spectral
  slope by fitting a straight line to each spectrum with the emission
  lines masked, and for all cases except the weakly varying Mrk~374,
  we found a significant anticorrelation between the mean flux and the
  spectral slope.  Detecting the ``bluer when brighter'' effect lends
  additional confidence to our relative flux calibration.

We also obtained six epochs of observations with the 2.3m telescope at
Wyoming Infrared Observatory (WIRO) and the WIRO Long Slit
Spectrograph.  The WIRO spectra were used to fill in gaps in the MDM
monitoring, and we matched the spectrograph configuration to that of
the MDM spectrograph as closely as possible.  This includes a 5\farcs0
slit at position angle 0$^\circ$ for all observations, and we used the
same extraction/sky apertures as for the MDM observations.  The
wavelength calibrations and spectral slopes of the WIRO data agree
well with the MDM observations, and we discuss the calibration of the
WIRO data to the MDM flux scale in \S2.5.1.

\subsubsection{Night-to-Night Flux Calibration}
In order to account for variable atmospheric extinction and seeing, we
employ the calibration algorithms introduced by \citet{Fausnaugh2017}.
This approach is similar to the older method of
\citet{vanGroningen1992}, but yields markedly better calibrations.  We
assume that the \oiii\ emission line is constant over the course of
our campaign, and we transform the observed spectra so that their
\oiii\ line profiles match those of the ``photometric'' nights (nights
with clear conditions and stable seeing).  We treat the WIRO and MDM
spectra separately and inter-calibrate the two flux scales below
(\S2.5.1).

\citet{Fausnaugh2017} discusses the details of our implementation and
provides a {\tt python} package ({\tt
  mapspec}\footnote{\url{https://github.com/mmfausnaugh/mapspec}}) to
build and apply a rescaling model to time-series spectra.  For
completeness, we briefly outline the procedure here:
\begin{enumerate}[i.]
\item First, we collected the spectra taken on photometric nights (as
  judged by the observers onsite) and estimated their \oiii\ line
  fluxes.  The line fluxes were measured by subtracting a linearly
  interpolated estimate of the local continuum underneath the line and
  then integrating the remaining flux using Simpson's method.  We
  provide the wavelength regions of the integration and the continuum
  fit in Tables \ref{tab:windows} and \ref{tab:con_windows}. We
  applied iterative 3$\sigma$ clipping to the line fluxes, where
  $\sigma$ is their root-mean-square (rms) scatter, in order to reject
  any outliers (due to slit losses or anomalies in the sky
  conditions).  We then averaged the remaining flux measurements to
  estimate the true line flux.  The measured \oiii\ line fluxes for
  each object are given in Table\,\ref{tab:targets}.  Table
  \ref{tab:targets} also gives the number of photometric epochs used
  to determine these fluxes for each AGN (we usually took three
  spectra per epoch).

\item We then combined the remaining photometric spectra into a
  reference spectrum using a noise-weighted average.  In this step,
  any residual wavelength shifts were removed by aligning the \oiii\
  line profiles using Markov Chain Monte Carlo (MCMC) methods---the
  spectra are shifted by the wavelength shift that minimizes the sum
  of the squares of residuals between the \oiii\ line profiles.
  Linear interpolation is used for wavelength shifts of fractional
  pixels.

\item Due to changes in seeing, spectrograph focus, and small guiding
  errors, the spectral resolution of each observation is slightly
  different.  To address this, we smooth the reference spectrum with a
  Gaussian kernel so that the \oiii\ linewidth matches the largest
  \oiii\ linewidth in the time series.  The smoothed reference
  spectrum will define the final resolution of the calibrated spectra.

\item The time-series spectra are then aligned to the reference by
  matching the \oiii\ line profiles, again in a least-squares sense
  using MCMC methods.  The differences in line profiles are modeled by
  a flux rescaling factor, a wavelength shift, and a smoothing kernel.
  After rescaling, we combine spectra from a single night using a
  noise-weighted average.
\end{enumerate}

\subsection{Imaging}
Our spectroscopic observations are supplemented with broad-band
imaging observations.  Contributing telescopes were the 0.7m at the
Crimean Astrophysical Observatory (CrAO), the 0.5m Centurian 18 at
Wise Observatory (WC18, \citealt{Brosch2008}), and the 0.9m at West
Mountain Observatory (WMO).  CrAO uses an AP7p CCD with a pixel scale
of 1\farcs76 and a $15' \times 15'$ field of view, WC18 uses a
STL-6303E CCD with a pixel scale of 1\farcs47 and a $75' \times 50'$
field of view, and WMO uses a Finger Lakes PL-3041-UV CCD with a pixel
scale of 0\farcs61 and a field of view of $21' \times 21'$.
Fountainwood Observatory (FWO) also provided observations of NGC\,4051
with a 0.4m telescope using an SBIG 8300M CCD.  The pixel scale of
this detector is 0\farcs35 and the field of view is $19' \times 15'$.\
All observations were taken with the Bessell {\it V}-band.

In addition, we obtained {\it ugriz} imaging with the LCO 1m network
\citep{Brown2013}, which consists of nine identical 1m telescopes at
four observatories spread around the globe.  These data were
originally acquired as part of LCO's AGN Key project
\citep{Valenti2015}.  The main goal is to search for continuum
reverberation signals, which we will pursue in a separate study
(Fausnaugh et al., in preparation).  However, 3C\,382 and Mrk\,374,
which are our faintest sources, had low variability amplitudes and
poorer S/Ns, so we included the LCO {\it g}-band data in the continuum
light curves of these objects.  Each LCO telescope has the same optic
system and detectors---at the time of the RM campaign, the detectors
were SBIGSTX-16803 cameras with a field of view of $16' \times 16'$
and a pixel scale of $0\farcs 23$.

We analyzed the imaging data using the image subtraction software
({\tt ISIS}) developed by \citet{Alard1998}.  Images were first
uploaded to a central repository and vetted by eye for obvious
reduction errors or poor observing conditions.  We then registered the
images to a common coordinate system and constructed a high S/N
reference frame by combining the best-seeing and lowest-background
images.  When combining, {\tt ISIS} adjusts the images to a common
seeing by convolving the point-spread function (PSF) of each image
with a spatially variable kernel.  Finally, we subtracted the
reference frame from each image, again allowing {\tt ISIS} to match
the PSFs using its convolution routine. Reference images and
subtractions for each telescope/filter/detection system were
constructed separately---we discuss combining the photometric
measurements in \S2.5.2.

\subsection{Mean and rms spectra}
Figures \ref{fig:mcg0811}--\ref{fig:mrk374} show the noise-weighted
mean spectrum
\begin{align}
  \overline F(\lambda) =
  \frac{\sum_{i=1}^{N_t}F(\lambda,t_i)/\sigma^2(\lambda,t_i)}{\sum_{i=1}^{N_t}1/\sigma^2(\lambda,t_i)}
\end{align}
for each object using the MDM observations, where $F(\lambda,t_i)$ is
the flux density at epoch $t_i$ and $\sigma(\lambda,t_i)$ is its
uncertainty.  Figures \ref{fig:mcg0811}--\ref{fig:mrk374} also show
root-mean-square (rms) residual spectra, defined as
\begin{align}
  \sigma_{\rm rms}(\lambda) = \sqrt{\frac{1}{N_t -1} \sum_{i=1}^{N_t}
    \left[F(\lambda,t_i) - \overline F (\lambda) \right]^2}\label{equ:rms}.
\end{align}
By the Wiener-Khinchin theorem, this statistic is proportional to the
integrated variability power at each wavelength, so $\sigma_{\rm rms}$
is free of constant contaminants such as host-galaxy and narrow
emission line flux.  However, the total variability power contains
contributions from both intrinsic variations and from statistical
fluctuations/measurement uncertainties.  In order to separate these
components, we use a maximum-likelihood method
(cf. \citealt{Park2012a, Barth2015, DeRosa2015}).  We solve for the
intrinsic variability $\sigma_{\rm var}(\lambda)$ that minimizes the
negative log-likelihood

\begin{align}
  -2 \ln \mathcal{L} = 
  \sum_{i=1}^{N_t} \frac{\left[ F(\lambda,t_i) -  \hat F(\lambda)\right]^2}
                      {\sigma^2(\lambda,t_i) +  \sigma_{\rm var}^2(\lambda)}  \nonumber\\
+   \sum_{i=1}^{N_t} \ln \left[\sigma^2(\lambda,t_i) + \sigma_{\rm var}^2(\lambda) \right]
\label{equ:rms_opt}
\end{align}
where $\hat F(\lambda)$ is the ``optimal average'' weighted by
$\sigma^2(t_i) + \sigma^2_{\rm var}$.  We self-consistently fit for
$\hat F(\lambda)$ while solving for $\sigma_{\rm var}(\lambda)$, and
we show the estimate of $\sigma_{\rm var}(\lambda)$ with the red lines
in Figures \ref{fig:mcg0811}--\ref{fig:mrk374}.  In the limit that
$\sigma(\lambda,t_i)\rightarrow 0$, it is clear that $\sigma_{\rm var}
$ is equivalent to $ \sigma_{\rm rms}$.  For high S/N data such as
these, $\sigma_{\rm var}(\lambda)$ is nearly equal to
$\left[\sigma_{\rm rms}^2(\lambda) - \overline \sigma^2(\lambda)
\right]^{1/2}$, where $\overline \sigma^2(\lambda)$ is the average of the
squared measurement uncertainties across the time-series:
\begin{align}
\overline \sigma^2(\lambda) = \frac{1}{N}\sum_{i=1}^{N_t}\sigma^2(\lambda,t_i).
\end{align}
The overall effect is to reduce the squared amplitude of the
variability spectrum by the mean squared measurement uncertainty---in
all objects except for Mrk\,374, this effect is negligible.

\subsection{Light curves}
\floattable
\begin{deluxetable}{lrrrrrr}
\tablewidth{0pt}
\tablecaption{Observed-frame Integration Windows \label{tab:windows}}
\tablehead{\colhead{Object} & \colhead{5100\,\AA} & \colhead{H$\beta$} & \colhead{H$\gamma$} & \colhead{He{\sc ii}$\lambda$4686 } & \colhead{[O{\sc iii}]$\lambda$5007 } & \colhead{[O{\sc iii}]$\lambda$4959 }\\
&\colhead{(\AA)} &\colhead{(\AA)}&\colhead{(\AA)}&\colhead{(\AA)}&\colhead{(\AA)}&\colhead{(\AA)}}
\startdata
MCG+08-11-011 & 5190--5230 & 4890--5040 & 4375--4485 & 4650--4890 & 5085--5130 & 5040--5075 \\
NGC 2617 & 5170--5200 & 4835--5050 & 4310--4520 & 4620--4835 & 5055--5093 & 5010--5040 \\
NGC 4051 & 5115--5145 & 4835--4920 & 4315--4390 & 4610--4740 & 5000--5045 & 4955--4977 \\
3C 382 & 5380--5400 & 5000--5270 & 4425--4745 & 4795--4930 & 5275--5330 & 5228--5257 \\
Mrk 374 & 5315--5350 & 4995--5140 & 4490--4580 & 4765--4915 & 5205--5245 & 5160--5187 \\
\enddata
\end{deluxetable}

\floattable
\begin{deluxetable}{llrrrrr}
\tablewidth{0pt}
\tablecaption{Observed-frame Continuum Fitting Windows \label{tab:con_windows}}
\tablehead{\colhead{Object} & \colhead{Line Side} & \colhead{H$\beta$} & \colhead{H$\gamma$} & \colhead{He{\sc ii}$\lambda$4686 } & \colhead{[O{\sc iii}]$\lambda$5007 } & \colhead{[O{\sc iii}]$\lambda$4959 }\\
& &\colhead{(\AA)}&\colhead{(\AA)}&\colhead{(\AA)}&\colhead{(\AA)}&\colhead{(\AA)}
}
\startdata
MCG+08-11-011 & Blue & 4860--4890 & 4360--4375 & 4620--4650 & 5075--5085 & 5030--5040 \\
 & Red & 5130--5150 & 4485--4500 & 4860--4880 & 5130--5150 & 5075--5085 \\
\hline
NGC 2617 & Blue & 4820--4835 & 4300--4310 & 4585--4620 & 5050--5055 & 5000--5010 \\
 & Red & 5110--5150 & 4520--4535 & 4820--4835 & 5093--5098 & 5040--5050 \\
\hline
NGC 4051 & Blue & 4800--4835 & 4300--4315 & 4605--4615 & 4990--5000 & 4945--4955 \\
 & Red & 4920--4950 & 4390--4400 & 4740--4775 & 5045--5055 & 4977--4990 \\
\hline
3C382 & Blue & 4975--5000 & 4410--4425 & 4785--4795 & 5265--5275 & 5218--5228 \\
 & Red & 5385--5425 & 4745--4760 & 4930--4940 & 5330--5340 & 5257--5271 \\
\hline
Mrk 374 & Blue & 4970--4990 & 4455--4490 & 4690--4765 & 5190--5205 & 5150--5160 \\
 & Red & 5140--5160 & 4580--4600 & 4915--5000 & 5245--5255 & 5187--5195 \\
\enddata
\end{deluxetable}

\floattable
\begin{deluxetable}{llrrcrrrrr}
\tablewidth{0pt}
\tablecaption{Light-curve Properties \label{tab:lc_prop}}
\tablehead{
\colhead{Object} & \colhead{Light curve} & \colhead{$N_{t}$} & \colhead{$\Delta t_{\rm med}$} & \colhead{Uncertainty} & \colhead{$\bar F$} & \colhead{$\langle {\rm S/N} \rangle$} & \colhead{$F_{\rm var}$} & \colhead{${\rm (S/N)_{var}}$}&\colhead{$r_{\rm max}$}\\
& & &\colhead{(days)} &\colhead{Rescaling Factor} & & & &\\
\colhead{(1)}&\colhead{(2)} &\colhead{(3)}&\colhead{(4)}&\colhead{(5)}&\colhead{(6)}&\colhead{(7)}&\colhead{(8)}&\colhead{(9)}&\colhead{(10)}
}
\startdata
MCG+08-11-011 & 5100\,\AA & 190 & 0.59 & 1.51 & 4.49 & 71.2 & 0.10 & 67.4 & $\equiv 1$ \\
 & H$\beta$ & 86 & 1.01 & 1.58 & 3.79 & 103.1 & 0.07 & 47.8 & $ 0.90\pm  0.01$ \\
 & H$\gamma$ & 82 & 1.01 & 1.52 & 1.29 & 34.2 & 0.09 & 19.2 & $ 0.84\pm  0.03$ \\
 & He{\sc ii}$\lambda$4686 & 86 & 1.01 & 1.46 & 0.31 & 6.8 & 0.44 & 19.6 & $ 0.78\pm  0.04$ \\
\hline
NGC 2617 & 5100\,\AA & 161 & 0.92 & 1.81 & 5.17 & 57.2 & 0.09 & 44.4 & $\equiv 1$ \\
 & H$\beta$ & 61 & 1.01 & 1.91 & 3.31 & 39.5 & 0.10 & 21.1 & $ 0.61\pm  0.07$ \\
 & H$\gamma$ & 61 & 1.01 & 1.65 & 1.18 & 11.0 & 0.20 & 12.3 & $ 0.62\pm  0.07$ \\
 & He{\sc ii}$\lambda$4686 & 61 & 1.01 & 1.18 & 0.15 & 3.2 & 0.61 & 10.9 & $ 0.49\pm  0.08$ \\
\hline
NGC 4051 & 5100\,\AA & 270 & 0.47 & 1.00 & 12.90 & 191.7 & 0.02 & 49.7 & $\equiv 1$ \\
 & H$\beta$ & 107 & 0.96 & 3.42 & 3.14 & 45.7 & 0.09 & 31.1 & $ 0.59\pm  0.05$ \\
 & H$\gamma$ & 98 & 0.99 & 2.48 & 1.92 & 26.1 & 0.08 & 13.7 & $ 0.50\pm  0.06$ \\
 & He{\sc ii}$\lambda$4686 & 107 & 0.96 & 3.25 & 1.59 & 12.9 & 0.11 & 10.2 & $ 0.47\pm  0.07$ \\
\hline
3c382 & 5100\,\AA & 209 & 0.56 & 1.17 & 3.18 & 148.5 & 0.09 & 131.3 & $\equiv 1$ \\
 & H$\beta$ & 81 & 1.00 & 1.70 & 2.06 & 43.4 & 0.05 & 14.5 & $ 0.83\pm  0.25$ \\
 & H$\gamma$ & 81 & 1.00 & 1.66 & 0.60 & 8.1 & 0.07 & 3.6 & $ 0.59\pm  0.24$ \\
 & He{\sc ii}$\lambda$4686 & 81 & 1.00 & 1.90 & 0.13 & 3.2 & 0.19 & 3.9 & $ 0.61\pm  0.47$ \\
\hline
Mrk 374 & 5100\,\AA & 180 & 0.59 & 2.39 & 3.80 & 94.5 & 0.03 & 25.5 & $\equiv 1$ \\
 & H$\beta$ & 67 & 1.01 & 1.54 & 1.29 & 38.4 & 0.05 & 11.2 & $ 0.50\pm  0.07$ \\
 & H$\gamma$ & 67 & 1.01 & 1.79 & 0.56 & 18.7 & 0.04 & 3.8 & $ 0.42\pm  0.07$ \\
 & He{\sc ii}$\lambda$4686 & 67 & 1.01 & 1.41 & 0.18 & 8.2 & 0.25 & 11.9 & $ 0.56\pm  0.06$ \\
\enddata
\tablecomments{Column 3 gives the number of observations in each light
  curve. Column 4 gives the median cadence.  Column 5 gives the
  rescaling factor by which the statistical uncertainties are
  multiplied to account for additional systematic errors (see
  \S2.5.1).  Column 6 gives the mean flux level of each light curve.
  The rest-frame 5100\,\AA\ continuum light curves are in units of
  $10^{-15}$ erg cm$^{-2}$ s$^{-1}$ \AA$^{-1}$, and the emission line
  light curves are in units of $10^{-13}$ erg cm$^{-2}$ s$^{-1}$.
  Column 7 gives the mean signal-to-noise ratio $\langle {\rm S/N}
  \rangle$.  Column 8 gives the rms fractional variability defined in
  Equation \ref{equ:fracvar}.  Column 9 gives the approximate S/N at
  which we detect variability (see \S 2.5.3).  Column 10 gives the
  maximum value of the interpolated cross correlation function (see
  \S3.1).}
\end{deluxetable}

\subsubsection{Spectroscopic Light Curves}

We extracted spectroscopic light curves for the wavelength windows
listed in Table\,\ref{tab:windows} for each AGN. We chose
  these windows based on visual inspection of the variable line
  profiles in the $\sigma_{\rm var}(\lambda)$ spectra, with the main
  goal of capturing the strongest variations in the lines.  For
3C\,382, the component tentatively identified as \heii\ is
blue-shifted by almost 100\,\AA\ relative to the systematic redshift,
and if variable \heii\ has a similar profile as the Balmer lines in
this object, this component corresponds to the blue wing of the line.

The rest-frame 5100\,\AA\ continuum, which is relatively free of
emission/absorption lines, was estimated by averaging the flux density
in the listed wavelength region.  Emission-line fluxes were determined
in the same way as for the \oiii\ line.  First we subtracted a linear
least-squares fit to the local continuum underneath the emission line.
Wavelength regions for the continuum fits are given in Table
\ref{tab:con_windows}.  Then we integrated the remaining flux using
Simpson's method (we did not assume a functional form for the emission
line).  In cases where the broad H$\beta$ wing extends underneath
[O{\sc iii}]$\lambda$4959, we subtracted the narrow emission line
(again with a local linear approximation of the underlying flux)
before integrating the broad line.  We did not attempt to separate the
narrow components of H$\beta$ and H$\gamma$ from the broad components.
These narrow components act as constant flux-offsets for the light
curves.  

The continuum estimates can lead to significant systematic
  uncertainties, because the continuum-fitting windows may be
  contaminated by broad-line wing emission, and the local linearly
  interpolated continuum may leave residual continuum flux to be
  included in the line profile.  Both of these effects can introduce
  spurious correlations between the continuum and line light curves,
  which may biased the final lag estimates.  Because we use the
  $\sigma_{\rm var}(\lambda)$ spectra to select the line and continuum
  windows, variability in the line wings probably does not have a
  large impact on our results, and we have found the the resulting
  light curves (and their lags) are robust to five to ten angstrom
  changes in the continuum and line windows.  Larger shifts,
  especially as the continuum fitting windows move further from the
  lines, can result in significantly different lags (of order three
  times the statistical uncertainties).  Full spectral decompositions
  may be able to address this issue in future studies (see
  \citealt{Barth2015} for a detailed discussion).  We discuss these
  systematic uncertainties further in \S4.  

After we extracted line fluxes from the WIRO and MDM spectra, we
combined the measurements by forcing the light curves to be on the
same flux scale.  We used the mean MDM \oiii\ line to define this
scale, and multiplied the WIRO line fluxes so that the mean value
matched that of MDM.  A more sophisticated inter-calibration
  model would include an additive offset, to account for different
  amounts of host-galaxy starlight in the MDM and WIRO spectra.
  However, with the limited amount of WIRO data, additional
  calibration parameters cannot be well-constrained, and we found the
  simple multiplicative approach to be adequate.  The required
rescaling factors were 1.21 for MCG+08-11-011, 1.14 for NGC\,4051,
1.09 for 3C\,382, and 1.73 for Mrk\,374.  Weather at WIRO prevented
observations of NGC\,2617.

The statistical uncertainty on the continuum flux was estimated from
the standard deviation within the wavelength region,
\begin{align}
  \sigma(t_j) = \sqrt{  \frac{1}{N_{\lambda}-1} \sum_{i = 1}^{N_{\lambda}}
    \left[ F(\lambda_i,t_j) - \overline F(t_j)\right]^2},
\end{align}
where $\overline F(t_j)$ is the evenly-weighted average flux density at
epoch $t_j$.  Uncertainties on the line light curves were estimated
using a Monte Carlo approach: we perturbed the observed spectrum with
random deviates scaled to the uncertainty at each wavelength,
subtracted a new estimate of the underlying continuum (and the narrow
[O{\sc iii}]\,$\lambda4959$ line when appropriate), and re-integrated
the line flux.  The deviates were drawn from the multivariate normal
distribution defined by the covariance matrix of the rescaled
spectrum---these covariances can affect the statistical uncertainty by
a factor of two or more (see \citealt{Fausnaugh2017} for more
details).  We repeated this procedure $10^3$ times and took the
central 68\% confidence interval of the output flux distributions as
an estimate of the statistical uncertainty.

\begin{figure*}
\includegraphics[width=\textwidth]{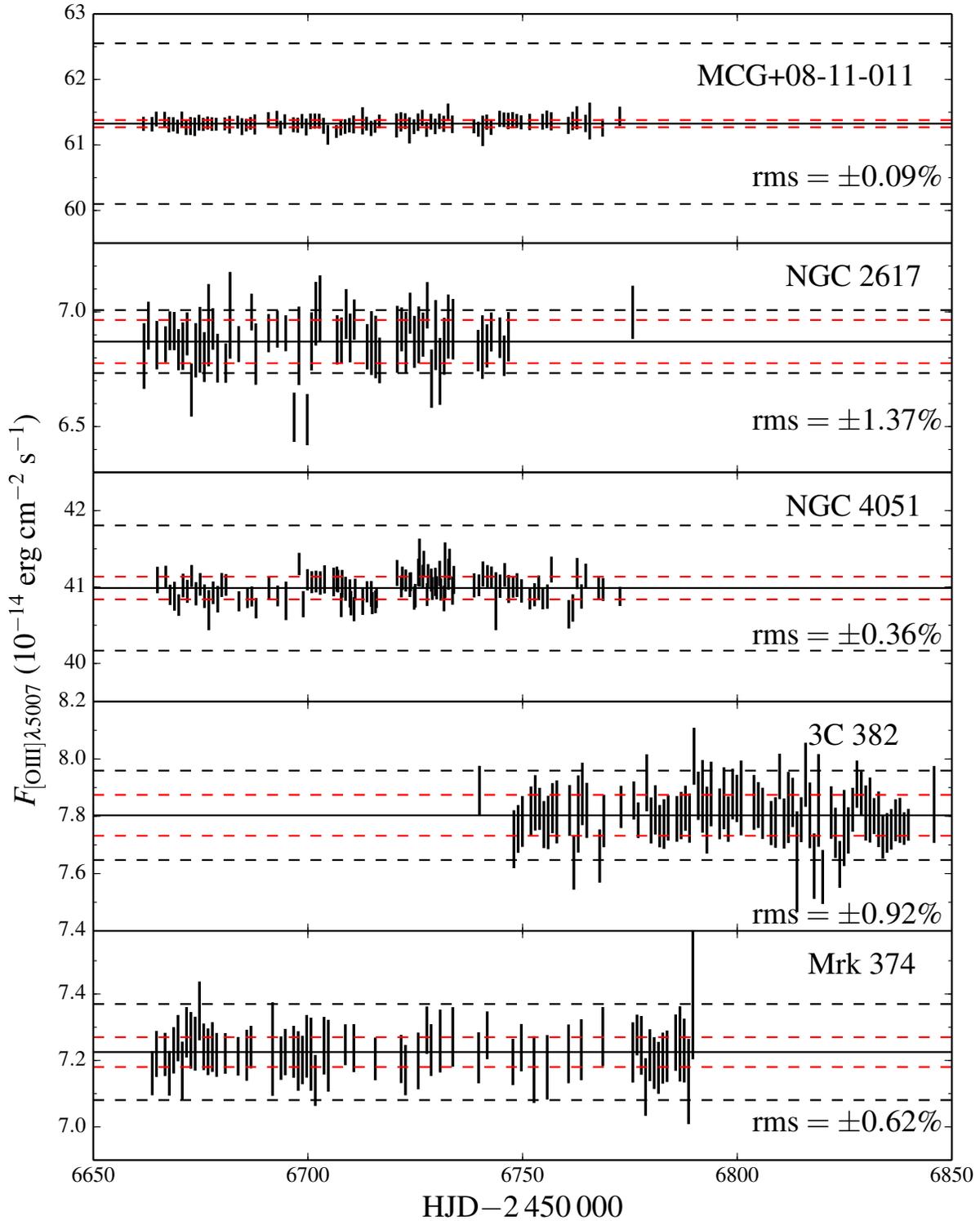}
\caption{Narrow \oiii\ line light curves.  The scatter of these light
  curves represents the precision of our final flux calibration.  The
  dashed black lines show 2\% variations around the mean, the dashed
  red lines show the rms scatter of the data, which range from 0.1\%
  (MCG+08-11-011) to 1.4\% (NGC\,2617).  Outliers are discussed in \S
  2.5.1.  \label{fig:oiii}}
\end{figure*}

Because the integrated \oiii\ line flux is not explicitly forced to be
equal from night to night, the scatter of the \oiii\ line light curve
serves as an estimate of our calibration uncertainty
\citep{Barth2015}.  We extracted narrow \oiii\ line light curves in
the same way as for the broad lines, and the results are shown in
Figure~\ref{fig:oiii}.  Several points are noticeably below the means
of their light curves, particularly for NGC\,2617 and 3C\,382.
These observations were taken in poor weather, and display
  significant scatter between the individual rescaled exposures prior
  to averaging.  This suggests variable amounts of flux-losses between
  the AGN and extended \oiii/host-galaxy, due to variable seeing and
  large guiding errors that move the object in the slit.  Although the
  rescaling model from \S2.2.2 cannot correct this issue, the offsets
  of these points are not very large compared to the statistical
  uncertainties (no more than 3.1$\sigma$), and we opt to include them
  in the analysis.  Since the effect due to spatially extended \oiii\
  emission is relatively small even in very poor conditions, it will
  be unimportant in good conditions.  

The fractional standard deviations of the narrow line light curves are
given in Table~\ref{tab:targets} and range between 0.1\% and 1.4\%.
These values only represent our ability to correct for
  extrinsic variations (such as weather conditions) in the observed
  spectra.  Additional systematic uncertainties dominate the
  epoch-to-epoch uncertainties of the light curves, including (but not
  limited to) the nightly sensitivity functions, continuum
  subtraction, and additional spectral components such as Fe{\sc ii}
  emission.  The latter two issues are especially problematic for the
\heii\ light curves.

To account for these systematics, we rescaled the light curve
uncertainties so that they approximate the observed flux variations
from night to night.  We selected three adjacent points $F(t_{j -1})$,
$F(t_j)$, and $F(t_{j +1})$, linearly interpolated between
$F(t_{j-1})$ and $F(t_{j+1})$, and measure $\Delta = [F(t_j) -
I(t_j)]/\sigma(t_{j})$ where $I(t_j)$ is the interpolated value at
$t_j$ and $\sigma(t_j)$ is the statistical uncertainty on $F(t_j)$.
The deviate $\Delta$ therefore measures the departure of the light
curve from a simple linear model. We calculated $\Delta$ for $j = 2$
to $N_t-1$ (i.e, ignoring the first and last points), and we
multiplied the statistical uncertainties $\sigma(t_j)$ by the mean
absolute deviation (MAD) $\overline{ |\Delta|}$.  We also imposed a a
minimum value of 1.0 on these rescaling factors.  Inspection
  of the distribution of $\Delta$ shows that the residuals are
  reasonably (but not perfectly) represented by a Gaussian with a
  similar MAD value.
  This method ensures that the uncertainties account for any
  systematics that the rescaling model cannot capture.  We have
ignored the uncertainty in the interpolation $I(t_j)$, so our method
slightly overestimates the required rescaling factors.  Monte Carlo
simulations may be able to assess the importance of uncertainty in
$I(t_j)$ for future work.  The rescaling factors are given in
Table\,\ref{tab:lc_prop} and are fairly small, generally running
between 1.0 and 2.0, with a mean of 1.8 and a maximum of 3.42 for the
H$\beta$ light curve in NGC\,4051.  NGC\,4051 has the largest
rescaling factors overall, which may be due to real short time-scale
variability that departs from our simple linear model
\citep{Denney2010}.  We therefore also experimented with using the
unscaled light curve uncertainties in our time-series analysis (\S3)
for this object.  We found that our results do not sensitively depend
on the scale of the uncertainties, although our Bayesian lag analysis
(\S3.2) indicates that the unscaled uncertainties are probably
underestimated.

\subsubsection{Broad-Band Light Curves}

Differential photometric light curves were extracted from the
subtracted broad-band images using {\tt ISIS}'s built-in photometry
package.  The software performs PSF photometry by fitting a model to
the reference frame PSF and convolving this model with the kernel that
was fitted during image subtraction.  Because this transformation
accounts for variable seeing, while the image subtraction has removed
sources of constant flux, the output light curves cleanly isolate
intrinsic variations of the AGN from contaminants such as host-galaxy
starlight and seeing-dependent aperture effects.  Any other
  constant systematic errors are also automatically subtracted out of
  the differential light curves.   However, {\tt ISIS} accounts for
only the local Poisson uncertainty from photon-counting, while there
are also systematic errors from imperfect subtractions (e.g.,
\citealt{Hartman2004}).  We addressed this problem in the same way as
\citet{Fausnaugh2016}.  We inspected the differential light curves of
comparison stars, and rescaled the uncertainties by a time dependent
factor to make the comparison star residuals consistent with a
constant model.  The reduced $\chi^2$ of the comparison star light
curves is therefore set to one, which requires an average error
rescaling factor of 1.0 to 5.0, depending on the object and the
telescope.  Since our targets are fairly bright, the formal
  ISIS uncertainties are very small and rescaling even by a factor of
  five results in uncertainties no greater than 3--6\%. See \S2.2 of
\citet{Fausnaugh2016} for more details.

We next calibrated the differential broad-band light curves to the
flux scale of the spectroscopic continuum light curve.  The
inter-calibration procedure solves for a maximum-likelihood shift and
rescaling factor for each differential light curve, forcing the {\it
  V}-band photometry to match the rest-frame 5100\,\AA\ continuum
flux.  The inter-calibration parameters account for the different
detector gains/bias levels, telescope throughputs, and (to
first-order) a correction for the wider bandpass and different
effective wavelengths of the broad-band filters compared to the
spectroscopic-continuum averaging window.  An advantage of
  this procedure is that it does not require accurate knowledge of the
  image zeropoints (or color corrections), which would otherwise limit
  the overall precision when combining data from different telescopes.
  The model also minimizes systematic errors that can result in strong
  correlations between measurements from the same telescope.

Because observations from various telescopes are never simultaneous,
it is necessary to interpolate the light curves when fitting the
inter-calibration parameters.  We followed \citet{Fausnaugh2016} and
modeled the time-series as a damped random walk (DRW), as implemented
by the {\tt JAVELIN} software \citep{Zu2011}.  Although recent studies
have shown that the power spectra of AGN light curves on short time
scales may be somewhat steeper than a DRW \citep{Edelson2014,
  Kasliwal2015}, \citet{Zu2013} found that the DRW is an adequate
description of the time scales considered here (see also
\citealt{Skielboe2015, Fausnaugh2016, Kozlowski2016a,
  Kozlowski2016b}).  Our interpolation scheme and fitting procedure
are identical to those described by \citet{Fausnaugh2016}.

\subsubsection{Light-curve Properties}

\begin{figure*}
\includegraphics[width=\textwidth]{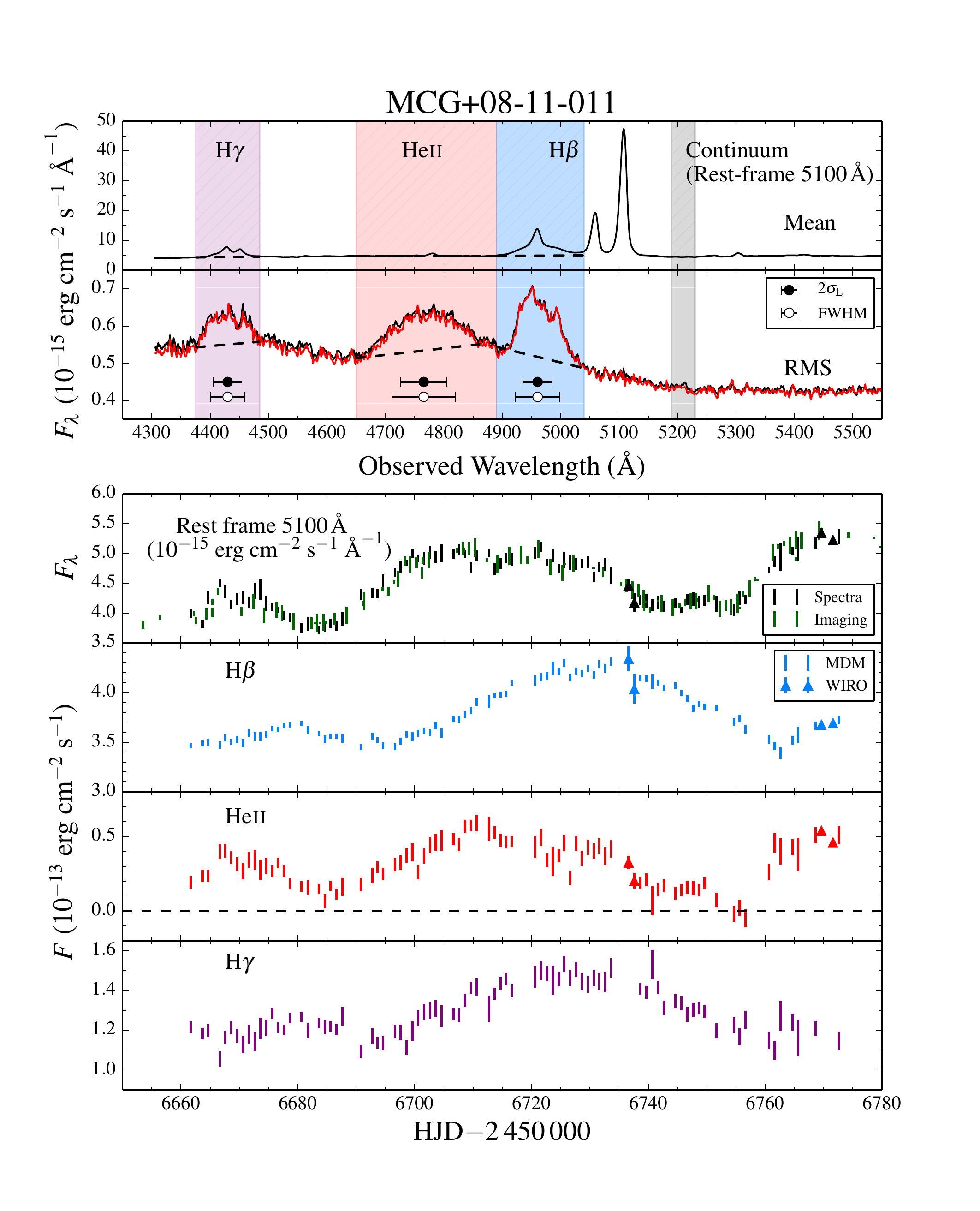}
\caption{Summary of the data for MCG+08-11-011.  Top Panel: Mean
  spectrum of the full time-series (top) and rms spectrum (bottom), as
  defined in \S2.4.  The black line is $\sigma_{\rm rms}(\lambda)$
  (Equation \ref{equ:rms}), and the red line is $\sigma_{\rm
    var}(\lambda)$, which includes a correction for measurement
  uncertainties (Equation \ref{equ:rms_opt}). The shaded regions show
  the windows from which continuum and line light curves were
  extracted.  The dashed lines show local linear fits to the continuum
  underlying the lines.  The errorbars show the rms linewidth
  ($\sigma_{\rm L}$) and full-width at half maximum (FWHM).  Bottom
  Panel: Light curves for the rest-frame 5100\,\AA\ continuum (imaging
  data is shown in green) and optical recombination
  lines. Over-subtraction of the continuum occasionally results in
  negative \heii\ fluxes, although RM measurements are only sensitive
  to the relative variations.\label{fig:mcg0811} }
\end{figure*}

\begin{figure*}
\includegraphics[width=\textwidth]{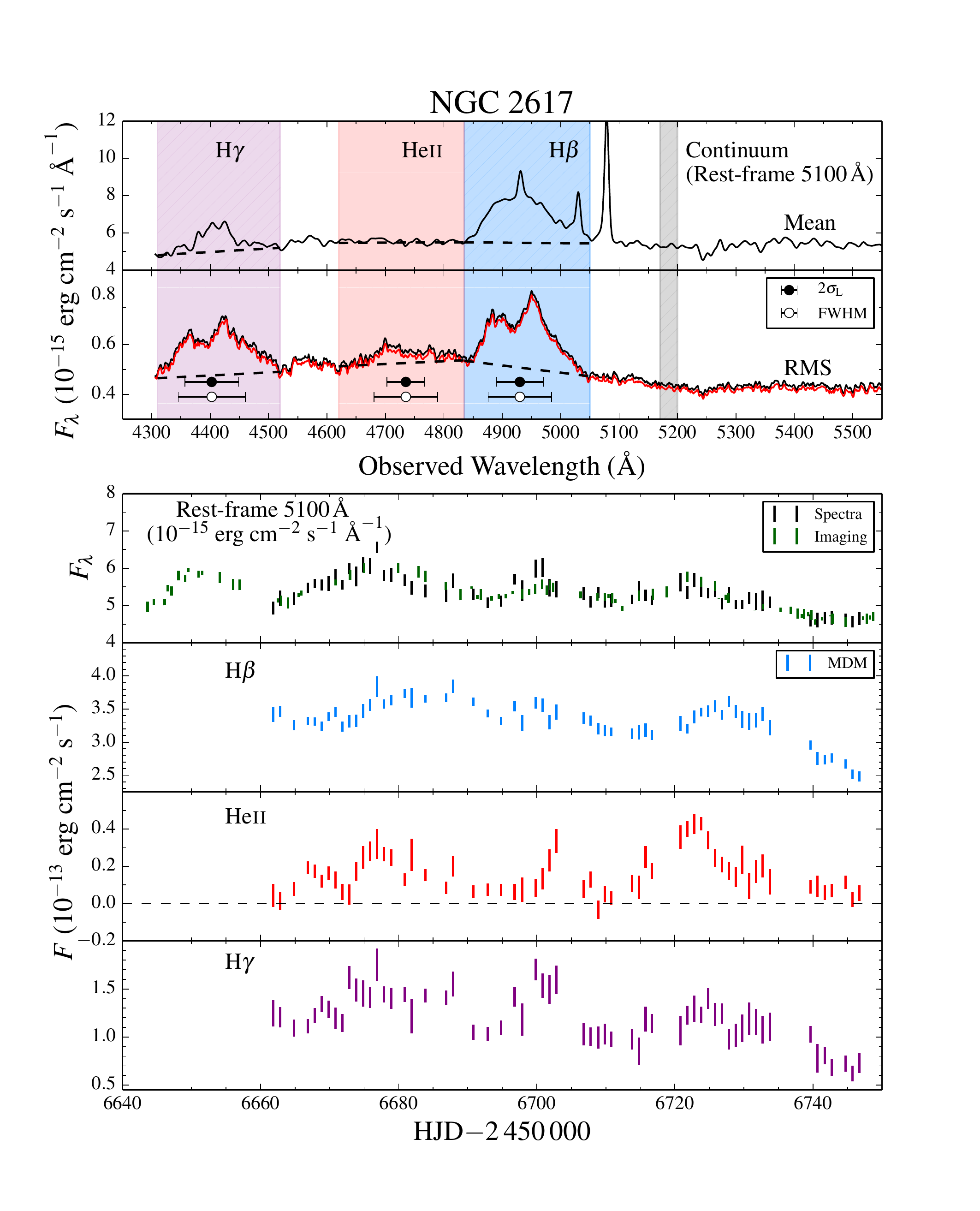}
\caption{Same as Figure\,\ref{fig:mcg0811} but for NGC
  2617.  \label{fig:n2617}}
\end{figure*}

\begin{figure*}
\includegraphics[width=\textwidth]{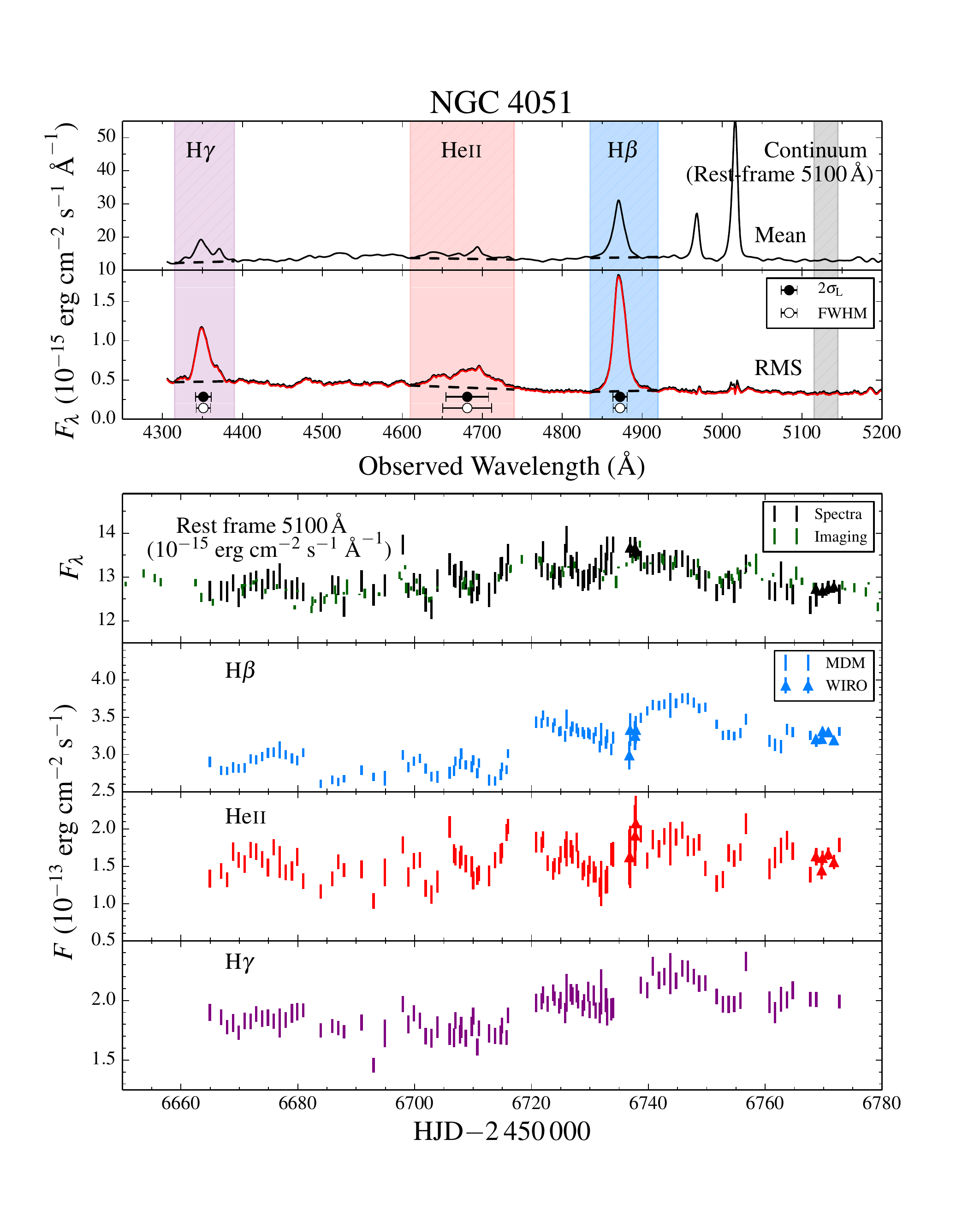}
\caption{Same as Figure\,\ref{fig:mcg0811} but for NGC
  4051.  \label{fig:n4051}}
\end{figure*}

\begin{figure*}
\includegraphics[width=\textwidth]{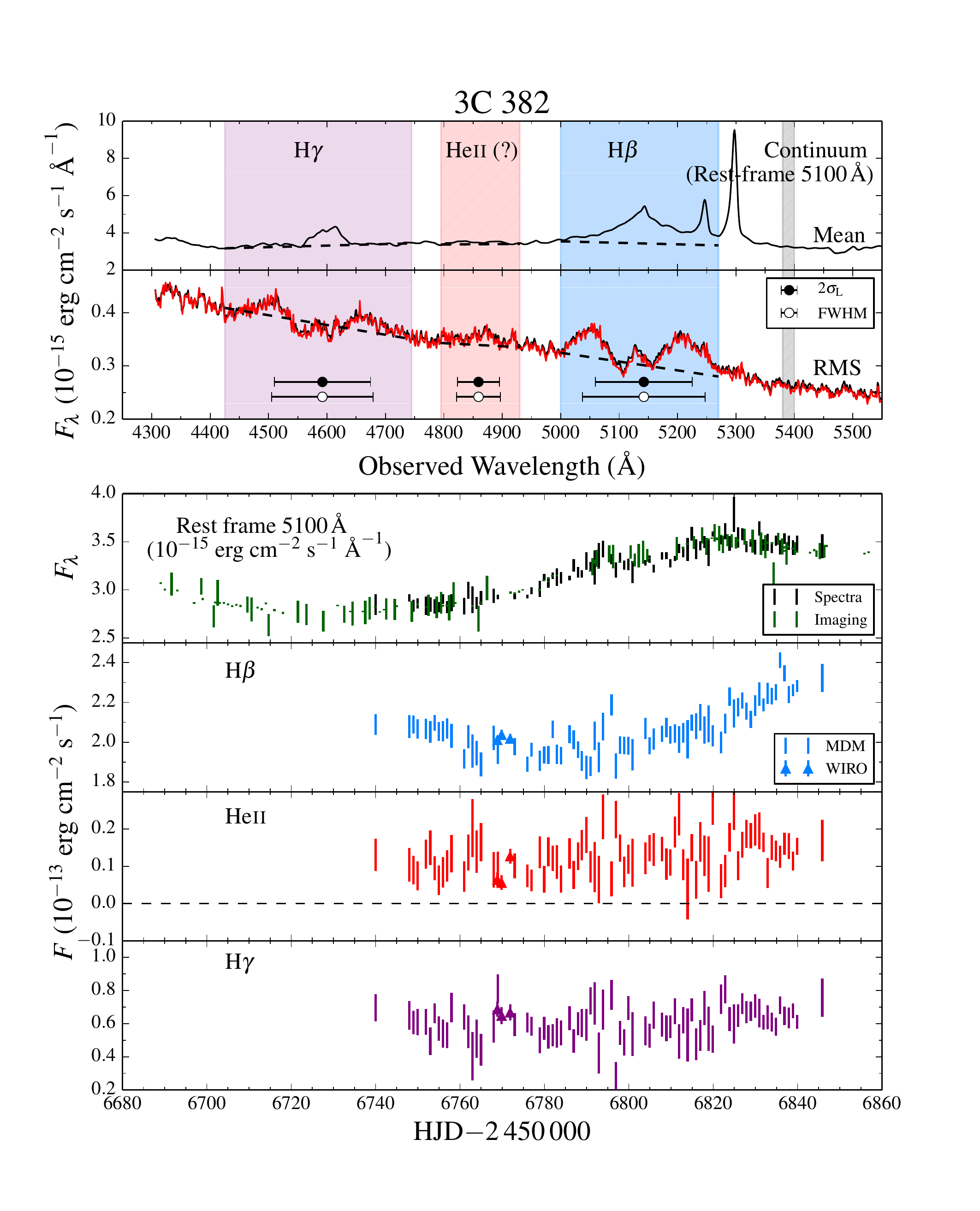}
\caption{Same as Figure\,\ref{fig:mcg0811} but for 3C\,382.  We do not
  use the H$\gamma$ or \heii\ light curves for the lag analysis in
  this study.\label{fig:3c382}}
\end{figure*}

\begin{figure*}
\includegraphics[width=\textwidth]{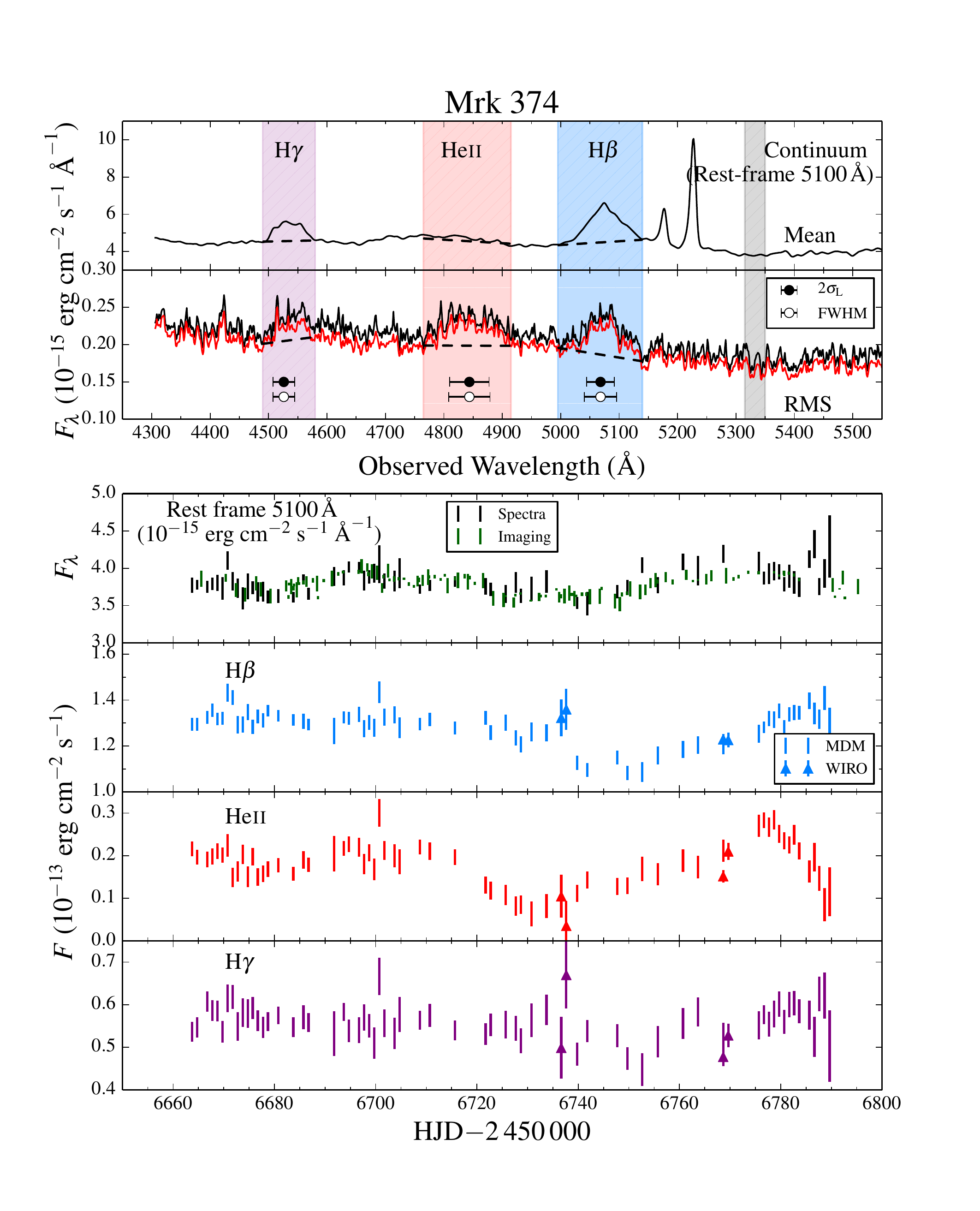}
\caption{Same as Figure\,\ref{fig:mcg0811} but for
  Mrk\,374.  \label{fig:mrk374}}
\end{figure*}

The final light curves are shown in Figures
\ref{fig:mcg0811}--\ref{fig:mrk374} and given in Tables 5--14.  We
characterize the statistical properties of the light curves in
Table\,\ref{tab:lc_prop}, reporting the median cadence, mean
flux-level, and average S/N.  We also measure the light curve
variability using a technique similar to our treatment of the
variability spectra $\sigma_{\rm var}(\lambda)$.  In the presence of
noise, it is necessary to separate the intrinsic variability from that
due to measurement errors.  We therefore define the intrinsic
variability of the light curves as $\sigma_{\rm var}$ and solve for it
by minimizing
\begin{align}
  -2 \ln \mathcal{L} = \sum_i^{N_t} \frac{\left[ F(t_i) - \hat
      F\right]^2}{\sigma^2(t_i) +  \sigma_{\rm var}^{2} } + \sum_i^{N_t} \ln
  \left[\sigma^2(t_i) + \sigma_{\rm var}^{ 2} \right],
\label{equ:fracvar}
\end{align}
where $F(t_i)$ is the flux at epoch $i$, $\sigma(t_i)$ is its
uncertainty, and $\hat F$ is the optimal average flux (weighted by
$\sigma^2(t_i) + \sigma^2_{\rm var}$).  For small measurement
uncertainties, the fractional variability $\sigma_{\rm var}/\hat F$
converges to the standard definition of the ``excess variance''
\citep{Rodriguez1997}
\begin{align}
  F_{\rm var} = \frac{1}{\overline F}\sqrt{\frac{1}{N -1} \sum_i^N
    \left[F(t_i) - \overline F  \right]^2 - \overline \sigma^2}
\end{align}
where $\overline \sigma$ is the time-averaged measurement uncertainty of
the light curve.  We therefore define $F_{\rm var} = \sigma_{\rm
  var}/\hat F$, and report these values in Table\,\ref{tab:lc_prop}.
These values are slightly underestimated, since $\hat F$ is not
corrected for constant components (such as host-galaxy starlight or
narrow line emission).  We also approximate the S/N of the variability
as
\begin{align} {\rm (S/N)_{var}} = \frac{\sigma_{\rm var}}{\overline
      \sigma \sqrt{ 2/N_{\rm obs}}}.
\end{align}
The $\sqrt{2/N_{\rm obs}}$ term enters because the variance of
$\overline \sigma$ is expected to approximately scale as that of a
reduced $\chi^2$ distribution.  However, this calculation assumes
uncorrelated uncertainties, and a full analysis requires treatment of
the red-noise properties of the light curve (see
\citealt{Vaughan2003}).

With the exception of the 3C\,382 H$\gamma$, the 3C\,382 \heii, and
the Mrk\,374 H$\gamma$ light curves, we detect variability in all of
the other emission lines at greater than $\sim \! 10\sigma$.  The
variability amplitudes of MCG+08-11-011 and NGC\,2617 are especially
strong ($F_{\rm var}\gtrsim10\%$).  For NGC\,4051, the continuum has
little fractional variability ($F_{\rm var} =2\%$), which may be
caused by a high fraction of host-galaxy starlight.  For
MCG+08-11-011, NGC\,2617, and NGC\,4051, the median cadence is near 1
day for all light curves, and the mean S/N usually ranges from several
tens to hundreds.  In fact, the S/N in the spectra is even higher,
reaching 100 to 300 per pixel in the continuum.  Combined with the
large variability amplitudes, it likely that we will be able to
construct velocity-delay maps and dynamical models for these objects
in future work.

\begin{deluxetable}{ccc}
\tablecaption{MCG+08-11-011 Continuum Light Curve}
\tablehead{
\colhead{HJD} & \colhead{$F_{\lambda}$} & \colhead{Telescope ID}\\
\colhead{(days)}&\colhead{($10^{-15}$ \,ergs\,s$^{-1}$\,cm$^{-2}$\,\AA$^{-1}$)}\\
\colhead{(1)}&\colhead{(2)}&\colhead{(3)}
}
\startdata
6639.5218 & $3.6279 \pm 0.0448$ & CrAO \\
6649.4750 & $3.6413 \pm 0.0845$ & CrAO \\
6653.4973 & $3.8005 \pm 0.0631$ & CrAO \\
6656.3942 & $3.9176 \pm 0.0425$ & CrAO \\
6661.6986 & $3.9866 \pm 0.0764$ & MDM \\
6662.2042 & $3.9837 \pm 0.0412$ & WC18 \\
\dots&\dots&\dots\\
\enddata
\tablecomments{Column 1 gives HJD $-$ 2\,450\,000 at
  mid-exposure. Column 2 give the continuum flux density and
  uncertainty.  Column 3 identifies the contributing telescope (see
  \S2.3).  A machine-readable version of this table is published in
  the electronic edition of this article. A portion is shown here for
  guidance regarding its form and content.}
\end{deluxetable}

\begin{deluxetable}{ccc}
\tablecaption{NGC 2617 Continuum Light Curve}
\tablehead{\colhead{HJD} & \colhead{$F_{\lambda}$} & \colhead{Telescope ID}\\
\colhead{(days)}&\colhead{($10^{-15}$\,ergs\,s$^{-1}$\,cm$^{-2}$\,\AA$^{-1}$)}\\
\colhead{(1)}&\colhead{(2)}&\colhead{(3)}
}
\startdata
6639.6747 & $4.6776 \pm 0.1089$ & CrAO \\
6643.6320 & $4.9717 \pm 0.1357$ & CrAO \\
6644.5145 & $5.0946 \pm 0.0871$ & CrAO \\
6646.0981 & $5.0904 \pm 0.0892$ & WC18 \\
6646.5287 & $5.4647 \pm 0.0885$ & CrAO \\
6647.0990 & $5.3930 \pm 0.1254$ & WC18 \\
\dots&\dots&\dots\\
\enddata
\tablecomments{Columns are the same as in Table 5.  A machine-readable
  version of this table is published in the electronic edition of this
  article. A portion is shown here for guidance regarding its form and
  content.}
\end{deluxetable}

\begin{deluxetable}{ccc}
\tablecaption{NGC 4051 Continuum Light Curve}
\tablehead{\colhead{HJD} & \colhead{$F_{\lambda}$} & \colhead{Telescope ID}\\
\colhead{(days)}&\colhead{($10^{-15}$\,ergs\,s$^{-1}$\,cm$^{-2}$\,\AA$^{-1}$)}\\
\colhead{(1)}&\colhead{(2)}&\colhead{(3)}
}
\startdata
6645.6113 & $12.7318 \pm 0.0624$ & WC18 \\
6646.6098 & $12.8314 \pm 0.0668$ & WC18 \\
6647.6275 & $13.0803 \pm 0.0683$ & WC18 \\
6648.5919 & $12.9598 \pm 0.0499$ & WC18 \\
6650.5178 & $12.8375 \pm 0.0550$ & WC18 \\
6653.5959 & $13.0814 \pm 0.0956$ & WC18 \\
\dots&\dots&\dots\\
\enddata
\tablecomments{Columns are the same as in Table 5.  A machine-readable
  version of this table is published in the electronic edition of this
  article. A portion is shown here for guidance regarding its form and
  content.}
\end{deluxetable}

\begin{deluxetable}{rrrcrrrrrrr}
\tablecaption{3C 382 Continuum Light Curve}
\tablehead{\colhead{HJD} & \colhead{$F_{\lambda}$} & \colhead{Telescope ID}\\
\colhead{(days)}&\colhead{($10^{-15}$\,ergs\,s$^{-1}$\,cm$^{-2}$\,\AA$^{-1}$)}\\
\colhead{(1)}&\colhead{(2)}&\colhead{(3)}
}
\startdata
6670.6411 & $3.2490 \pm 0.1390$ & WC18 \\
6689.0237 & $3.0685 \pm 0.0094$ & LCOGT1 \\
6690.0279 & $3.0018 \pm 0.0105$ & LCOGT1 \\
6691.6169 & $3.0619 \pm 0.1197$ & WC18 \\
6693.0025 & $2.9948 \pm 0.0106$ & LCOGT1 \\
6696.9897 & $2.8652 \pm 0.0094$ & LCOGT1 \\
\dots&\dots&\dots\\
\enddata
\tablecomments{Columns are the same as in Table 5.  A machine-readable
  version of this table is published in the electronic edition of this
  article. A portion is shown here for guidance regarding its form and
  content.}
\end{deluxetable}

\begin{deluxetable}{ccc}
\tablecaption{Mrk 374 Continuum Light Curve}
\tablehead{\colhead{HJD} & \colhead{$F_{\lambda}$} & \colhead{Telescope ID}\\
\colhead{(days)}&\colhead{($10^{-15}$\,ergs\,s$^{-1}$\,cm$^{-2}$\,\AA$^{-1}$)}\\
\colhead{(1)}&\colhead{(2)}&\colhead{(3)}
}
\startdata
6663.7432 & $3.7706 \pm 0.1030$ & MDM \\
6664.7221 & $3.7776 \pm 0.0649$ & MDM \\
6665.5588 & $3.8676 \pm 0.1114$ & WC18 \\
6666.7292 & $3.7496 \pm 0.0791$ & MDM \\
6667.7164 & $3.7902 \pm 0.0850$ & MDM \\
6668.7316 & $3.8045 \pm 0.1097$ & MDM \\
\dots&\dots&\dots\\
\enddata
\tablecomments{Columns are the same as in Table 5.  A machine-readable
  version of this table is published in the electronic edition of this
  article. A portion is shown here for guidance regarding its form and
  content.}
\end{deluxetable}

\floattable
\begin{deluxetable}{ccccc}
\tablecaption{MCG+08-11-011 Line Light Curves}
\tablehead{
\colhead{HJD} & \colhead{H$\beta$} & \colhead{H$\gamma$} & \colhead{He{\sc ii}} & \colhead{Telescope ID}\\
\colhead{(days)}&\multicolumn{3}{c}{($10^{-13}$\,erg\,s$^{-1}$\,cm$^{-2}$)} \\
\colhead{(1)}&\colhead{(2)}&\colhead{(3)}&\colhead{(4)}&\colhead{(5)}
}
\startdata
6661.6986 & $3.4660 \pm 0.0288$ & $0.0298 \pm 0.0298$ & $0.1923 \pm 0.0416$ & MDM \\
6663.6924 & $3.4891 \pm 0.0302$ & $0.0291 \pm 0.0291$ & $0.2341 \pm 0.0401$ & MDM \\
6664.6734 & $3.4978 \pm 0.0346$ & $0.0326 \pm 0.0326$ & $0.2352 \pm 0.0407$ & MDM \\
6666.6277 & $3.4723 \pm 0.0387$ & $0.0388 \pm 0.0388$ & $0.3973 \pm 0.0505$ & MDM \\
6667.6147 & $3.5429 \pm 0.0365$ & $0.0354 \pm 0.0354$ & $0.4047 \pm 0.0471$ & MDM \\
6668.6285 & $3.5031 \pm 0.0307$ & $0.0333 \pm 0.0333$ & $0.3580 \pm 0.0437$ & MDM \\
\dots&\dots&\dots&\dots&\dots\\
\enddata
\tablecomments{Column 1 gives HJD $-$ 2\,450\,000 at
  mid-exposure. Columns 2--4 give the line fluxes and their
  uncertainties.  Column 5 identifies the contributing telescope (see
  \S2.3). A machine-readable version of this table is published in the
  electronic edition of this article. A portion is shown here for
  guidance regarding its form and content.}
\end{deluxetable}

\floattable
\begin{deluxetable}{ccccc}
\tablecaption{NGC 2617 Line Light Curves}
\tablehead{
\colhead{HJD} & \colhead{H$\beta$} & \colhead{H$\gamma$} & \colhead{He{\sc ii}} & \colhead{Telescope ID}\\
\colhead{(days)}&\multicolumn{3}{c}{($10^{-13}$\,erg\,s$^{-1}$\,cm$^{-2}$)} \\
\colhead{(1)}&\colhead{(2)}&\colhead{(3)}&\colhead{(4)}&\colhead{(5)}
}
\startdata
6661.8270 & $3.4179 \pm 0.1151$ & $1.2480 \pm 0.1362$ & $0.0439 \pm 0.0614$ & MDM \\
6662.8456 & $3.4662 \pm 0.0848$ & $1.2049 \pm 0.1039$ & $0.0145 \pm 0.0470$ & MDM \\
6664.8143 & $3.2528 \pm 0.0756$ & $1.0938 \pm 0.0892$ & $0.0779 \pm 0.0370$ & MDM \\
6666.8417 & $3.3192 \pm 0.0635$ & $1.1110 \pm 0.0744$ & $0.1912 \pm 0.0346$ & MDM \\
6667.8446 & $3.3112 \pm 0.0621$ & $1.2235 \pm 0.0793$ & $0.1744 \pm 0.0363$ & MDM \\
6668.8507 & $3.2390 \pm 0.0690$ & $1.3410 \pm 0.0837$ & $0.1203 \pm 0.0344$ & MDM \\
\dots&\dots&\dots&\dots&\dots\\
\enddata
\tablecomments{Columns are the same as in Table 10.  A
  machine-readable version of this table is published in the
  electronic edition of this article. A portion is shown here for
  guidance regarding its form and content.}
\end{deluxetable}

\floattable
\begin{deluxetable}{ccccc}
\tablecaption{NGC 4051 Line Light Curves}
\tablehead{
\colhead{HJD} & \colhead{H$\beta$} & \colhead{H$\gamma$} & \colhead{He{\sc ii}} & \colhead{Telescope ID}\\
\colhead{(days)}&\multicolumn{3}{c}{($10^{-13}$\,erg\,s$^{-1}$\,cm$^{-2}$)}\\
\colhead{(1)}&\colhead{(2)}&\colhead{(3)}&\colhead{(4)}&\colhead{(5)}
}
\startdata
6664.9552 & $2.9038 \pm 0.0705$ & $0.0729 \pm 0.0729$ & $1.3355 \pm 0.1207$ & MDM \\
6666.8941 & $2.7895 \pm 0.0616$ & $0.0659 \pm 0.0659$ & $1.4384 \pm 0.1032$ & MDM \\
6667.8962 & $2.7808 \pm 0.0565$ & $0.0605 \pm 0.0605$ & $1.3202 \pm 0.0981$ & MDM \\
6668.9146 & $2.8279 \pm 0.0708$ & $0.0767 \pm 0.0767$ & $1.6972 \pm 0.1251$ & MDM \\
6669.9111 & $2.8152 \pm 0.0586$ & $0.0599 \pm 0.0599$ & $1.5860 \pm 0.1034$ & MDM \\
6670.9090 & $2.8217 \pm 0.0657$ & $0.0694 \pm 0.0694$ & $1.5306 \pm 0.1166$ & MDM \\
\dots&\dots&\dots&\dots&\dots\\
\enddata
\tablecomments{Columns are the same as in Table 10.  A
  machine-readable version of this table is published in the
  electronic edition of this article. A portion is shown here for
  guidance regarding its form and content.}
\end{deluxetable}

\floattable
\begin{deluxetable}{ccccc}
\tablecaption{3C 382 Line Light Curves}
\tablehead{
\colhead{HJD} & \colhead{H$\beta$} & \colhead{H$\gamma$} & \colhead{He{\sc ii}} & \colhead{Telescope ID}\\
\colhead{(days)}&\multicolumn{3}{c}{($10^{-13}$\,erg\,s$^{-1}$\,cm$^{-2}$)} \\
\colhead{(1)}&\colhead{(2)}&\colhead{(3)}&\colhead{(4)}&\colhead{(5)}
}
\startdata
6739.9650 & $2.0891 \pm 0.0519$ & $0.6969 \pm 0.0820$ & $0.1308 \pm 0.0429$ & MDM \\
6747.9747 & $2.0787 \pm 0.0581$ & $0.6508 \pm 0.0864$ & $0.1038 \pm 0.0452$ & MDM \\
6748.9640 & $2.0885 \pm 0.0447$ & $0.6067 \pm 0.0715$ & $0.0901 \pm 0.0388$ & MDM \\
6749.9571 & $2.0675 \pm 0.0487$ & $0.6094 \pm 0.0785$ & $0.0745 \pm 0.0392$ & MDM \\
6751.9471 & $2.0665 \pm 0.0520$ & $0.6116 \pm 0.0784$ & $0.1327 \pm 0.0392$ & MDM \\
6752.9466 & $2.0513 \pm 0.0550$ & $0.4939 \pm 0.0833$ & $0.1497 \pm 0.0464$ & MDM \\
\dots&\dots&\dots&\dots&\dots\\
\enddata
\tablecomments{Columns are the same as in Table 10.  A
  machine-readable version of this table is published in the
  electronic edition of this article. A portion is shown here for
  guidance regarding its form and content.}
\end{deluxetable}

\floattable
\begin{deluxetable}{ccccc}
\tablecaption{Mrk 374 Line Light Curves}
\tablehead{
\colhead{HJD} & \colhead{H$\beta$} & \colhead{H$\gamma$} & \colhead{He{\sc ii}} & \colhead{Telescope ID}\\
\colhead{(days)}&\multicolumn{3}{c}{($10^{-13}$\,erg\,s$^{-1}$\,cm$^{-2}$)} \\
\colhead{(1)}&\colhead{(2)}&\colhead{(3)}&\colhead{(4)}&\colhead{(5)}
}
\startdata
6663.7432 & $1.2948 \pm 0.0280$ & $0.5368 \pm 0.0232$ & $0.2157 \pm 0.0174$ & MDM \\
6664.7221 & $1.2944 \pm 0.0274$ & $0.5467 \pm 0.0238$ & $0.1965 \pm 0.0170$ & MDM \\
6666.7292 & $1.3247 \pm 0.0282$ & $0.6074 \pm 0.0243$ & $0.1911 \pm 0.0188$ & MDM \\
6667.7164 & $1.3559 \pm 0.0291$ & $0.5865 \pm 0.0245$ & $0.1980 \pm 0.0193$ & MDM \\
6668.7316 & $1.3172 \pm 0.0278$ & $0.5855 \pm 0.0245$ & $0.2111 \pm 0.0184$ & MDM \\
6669.7373 & $1.3207 \pm 0.0282$ & $0.5365 \pm 0.0254$ & $0.2015 \pm 0.0181$ & MDM \\
\dots&\dots&\dots&\dots&\dots\\
\enddata
\tablecomments{Columns are the same as in Table 5.  A machine-readable
  version of this table is published in the electronic edition of this
  article. A portion is shown here for guidance regarding its form and
  content.}
\end{deluxetable}

\section{Time-series Measurements}
We measure lags between continuum and line light curves using two
independent methods: traditional cross-correlation techniques and a
Bayesian analysis using the {\tt JAVELIN} software.
\floattable
\begin{deluxetable}{llrrrr}
\tablewidth{0pt}
\tablecaption{Rest-frame Line Lags \label{tab:linelags}}
\tablehead{
\colhead{Object} & \colhead{Line} & \colhead{$\tau_{\rm cent}$} & \colhead{$\tau_{\rm peak}$} & \colhead{$\tau_{\tt JAV}$} & \colhead{$\tau_{\rm multi}$} \\
& &\colhead{(days)}&\colhead{(days)}&\colhead{(days)}&\colhead{(days)}\\
\colhead{(1)}&\colhead{(2)}&\colhead{(3)}&\colhead{(4)}&\colhead{(5)}&\colhead{(6)}
}
\startdata
MCG+08-11-011 & H$\beta$ & $15.72^{+0.50} _{-0.52}$ & $15.02^{+1.86} _{-1.08}$ & $15.06^{+0.26} _{-0.28}$ & $14.98^{+0.34}_{-0.28}$ \\
 & H$\gamma$ & $13.14^{+1.12} _{-1.05}$ & $12.08^{+0.69} _{-0.98}$ & $11.92^{+0.44} _{-0.44}$ & $12.38^{+0.46}_{-0.49}$ \\
 & He{\sc ii}$\lambda$4686 & $1.88^{+0.58} _{-0.64}$ & $1.59^{+1.27} _{-0.98}$ & $1.24^{+0.36} _{-0.29}$ & $1.21^{+0.29}_{-0.33}$ \\
\hline
NGC 2617 & H$\beta$ & $4.32^{+1.10} _{-1.35}$ & $4.16^{+1.08} _{-1.68}$ & $6.22^{+0.51} _{-0.54}$ & $6.38^{+0.44}_{-0.50}$ \\
 & H$\gamma$ & $0.91^{+1.50} _{-1.08}$ & $0.61^{+1.28} _{-0.89}$ & $0.83^{+0.58} _{-0.59}$ & $0.81^{+0.59}_{-0.61}$ \\
 & He{\sc ii}$\lambda$4686 & $1.59^{+0.49} _{-0.69}$ & $1.79^{+0.20} _{-0.89}$ & $1.78^{+0.30} _{-0.35}$ & $1.75^{+0.34}_{-0.38}$ \\
\hline
NGC 4051 & H$\beta$ & $2.87^{+0.86} _{-1.33}$ & $2.42^{+1.00} _{-1.90}$ & $2.41^{+0.38} _{-0.46}$ & $2.24^{+0.39}_{-0.28}$ \\
 & H$\gamma$ & $2.82^{+1.82} _{-2.84}$ & $2.82^{+1.90} _{-3.09}$ & $4.87^{+0.28} _{-0.08}$ & $2.40^{+0.86}_{-0.73}$ \\
 & He{\sc ii}$\lambda$4686 & $0.27^{+0.33} _{-0.40}$ & $0.23^{+0.40} _{-0.40}$ & $0.06^{+0.17} _{-0.17}$ & $-0.03^{+0.16}_{-0.15}$ \\
\hline
3C382 & H$\beta$ & $40.49^{+8.02} _{-3.74}$ & $43.58^{+4.16} _{-3.50}$ & $52.07^{+3.18} _{-9.46}$ & \dots \\
\hline
Mrk 374 & H$\beta$ & $14.84^{+5.76} _{-3.30}$ & $14.81^{+5.85} _{-3.55}$ & $15.03^{+1.41} _{-1.26}$ & $13.73^{+1.06}_{-1.02}$ \\
 & H$\gamma$ & $12.31^{+9.82} _{-9.80}$ & $12.51^{+9.69} _{-11.80}$ & $15.44^{+3.26} _{-2.85}$ & $13.37^{+2.11}_{-2.08}$ \\
 & He{\sc ii}$\lambda$4686 & $-1.53^{+3.21} _{-5.79}$ & $-1.59^{+2.69} _{-5.85}$ & $-0.44^{+0.71} _{-0.68}$ & $-0.57^{+0.65}_{-0.64}$ \\
\enddata
\tablecomments{Column 3 and Column 4 give the centroids and peaks,
  respectively, of the interpolated cross correlation functions
  (ICCFs).  The uncertainties give the central 68\% confidence
  intervals of the ICCF distributions from the FR/RSS procedure (see
  \S3.1).  Column 5 gives the lag fit by {\tt JAVELIN}.  Column 6
  gives the same but using all light curves from a single object
  simultaneously.  The uncertainties give the central 68\% confidence
  intervals of the {\tt JAVELIN} posterior lag distributions.  All
  lags are relative to the 5100\,\AA\ continuum light curve and
  corrected to the rest-frame.  The uncertainties only represent the
  statistical errors---choices of continuum
  windows, detrending procedures, etc., introduce additional
  systematic uncertainties.}
\end{deluxetable}

\subsection{Cross-Correlation}
The cross-correlation procedure derives a lag from the centroid of the
interpolated cross-correlation function (ICCF, \citealt{Gaskell1987}),
as implemented by \citet{Peterson2004}.  For a given time delay, we
shift the abscissas of the first light curve, linearly interpolate the
second light curve to the new time coordinates, and calculate the
correlation coefficient $r_{cc}$ between all overlapping data points.
We then repeat this calculation but shift the second light curve by
the negative of the given time delay and interpolate the first light
curve.  The two values of $r_{cc}$ are averaged together, and the ICCF
is evaluated by repeating this procedure on a grid of time delays
spaced by 0.1 days.  All ICCFs are measured relative to the 5100\,\AA\
continuum light curve (inter-calibrated with the broad-band
measurements).  For each line light curve, the maximum value $r_{\rm
  max}$ of the ICCF is given in Table\,\ref{tab:lc_prop}.  The lag is
estimated with the ICCF centroid, defined as $\tau_{\rm cent} = \int
\tau r_{cc}(\tau)\,d\tau / \int r_{cc}(\tau)\,d\tau$ for values of $r_{cc} \geq 0.8 r_{\rm
  max}$.

We estimate the uncertainty on $\tau_{\rm cent}$ using the flux
randomization/random subset sampling (FR/RSS) method of
\citet{Peterson2004}.  This technique generates perturbed light curves
by randomly selecting (with replacement) a subset of the data from
both light curves and adjusting the fluxes by a Gaussian deviate
scaled to the measurement uncertainties.  The lag $\tau_{\rm cent}$ is
calculated for $10^3$ perturbations of the data, and its uncertainty
is estimated from the central 68\% confidence interval of the
resulting distribution.  The ICCF and centroid distributions are shown
in Figure\,\ref{fig:linelags} for all objects and line light curves,
and Table\,\ref{tab:linelags} gives the median values and central 68\%
confidence intervals of these distributions.  For completeness, we
also report in Table\,\ref{tab:linelags} the lag $\tau_{\rm peak}$
that corresponds to $r_{\rm max}$. Note that these lags have been
corrected to the rest frame of the source.  For 3C\,382, we do not
find meaningful centroids in the ICCFs of the H$\gamma$ and \heii\
light curves.  This is because of the width of the autocorrelation
function of the continuum and its poor correlation with the line light
curves.  We therefore do not include these lines for the rest of the
ICCF analysis.

Long-term trends in the light curves can bias the resulting ICCF due
to red-noise leakage \citep{Welsh1999}.  We therefore experimented
with detrending the light curves and/or restricting the baseline over
which to calculate the ICCF.  For MCG+08-11-011 these experiments had
no effect, while for Mrk\,374 and 3C\,382 they eliminated any lag
signal in the data.  For NGC\,2617, we found that restricting the data
to 6\,620\,$<$\,HJD\,$-$\,2\,450\,000\,$<$\,6\,730 improved the ICCF
by narrowing the central peak, as shown in the top four panels of
Figure \ref{fig:detrend}.  However, this restriction changed the ICCF
centroid by only 0.01 days, a negligible amount.  For NGC\,2617, the
peaks in the H$\gamma$ and \heii\ ICCFs at $\pm 25$ days are also
obvious aliases, so we only report the lag based on the peak near 0
days.  For NGC\,4051, we found that detrending the continuum and line
light curves with a second-order polynomial improves the ICCF, as
shown in the bottom four panels of Figure \ref{fig:detrend}.  The
long-term continuum trend is very weak, but there is a strong positive
trend in the line light curves that is dominated by the linear term.
Subtracting this linear trend decreases the median of the centroid
distribution from 4.92 days to 2.56 days, a change of 1.5$\sigma$.  We
adopt the smaller lag because of the quality of the detrended ICCF,
and our Bayesian method (described below) finds a lag consistent with
this smaller value.

\subsection{\tt JAVELIN}
We also investigated the line lags using a Bayesian approach, as
implemented by the {\tt JAVELIN} software \citep{Zu2011}.  {\tt
  JAVELIN} explicitly models the reverberating light curves and
corresponding transfer functions so as to find a posterior probability
distribution of lags.  We have already discussed {\tt JAVELIN}'s
assumption that light curves are reasonably characterized by a DRW
(\S2.5.2).  {\tt JAVELIN} also assumes that the transfer function is a
simple top-hat that can be parameterized by a width, an amplitude, and
a mean time delay.  This assumption is not very restrictive, since it
is difficult to distinguish among transfer functions in the presence
of noise \citep{Rybicki1994, Zu2011} and a top-hat is broadly
consistent with expectations for physically-plausible BLR geometries
(e.g., disks or spherical shells).

\begin{figure*}
\centering
\includegraphics[width=0.9\textwidth]{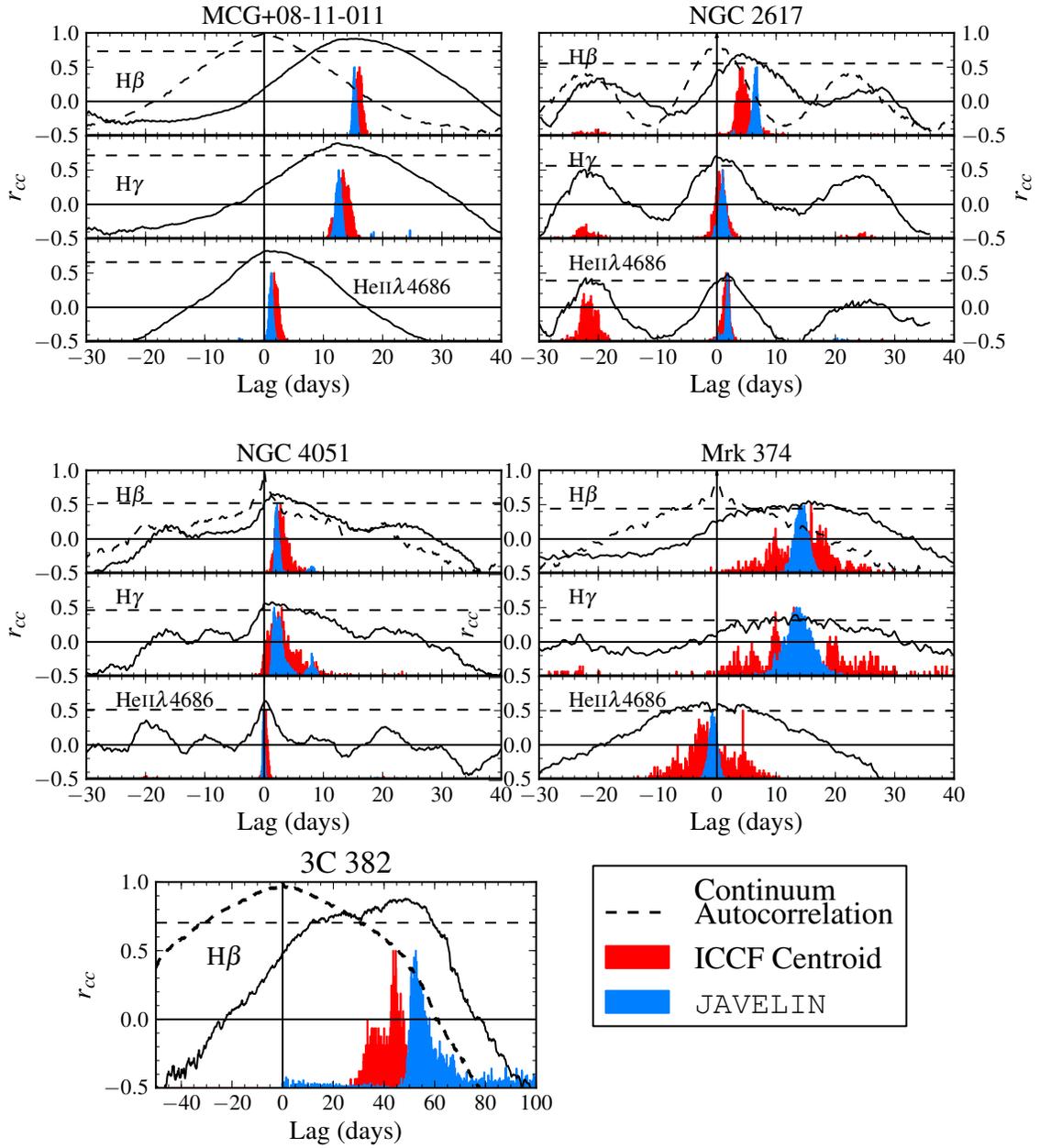}
\caption{Lags for all emission lines in all objects.  The solid lines
  show the ICCFs and the dashed lines show the autocorrelation
  function of the continuum.  The red histograms show the ICCF
  centroid distributions for $\tau_{\rm cent}$, and the blue
  histograms show the posterior {\tt JAVELIN} lag distributions for
  $\tau_{\rm multi}$.  The histograms are normalized by dividing by
  their maximum values.  The horizontal dashed lines show $0.8r_{\rm
    max}$, used to calculate $\tau_{\rm cent}$.  \label{fig:linelags}}
\end{figure*}

\begin{figure*}
\centering
\includegraphics[width=0.9\textwidth]{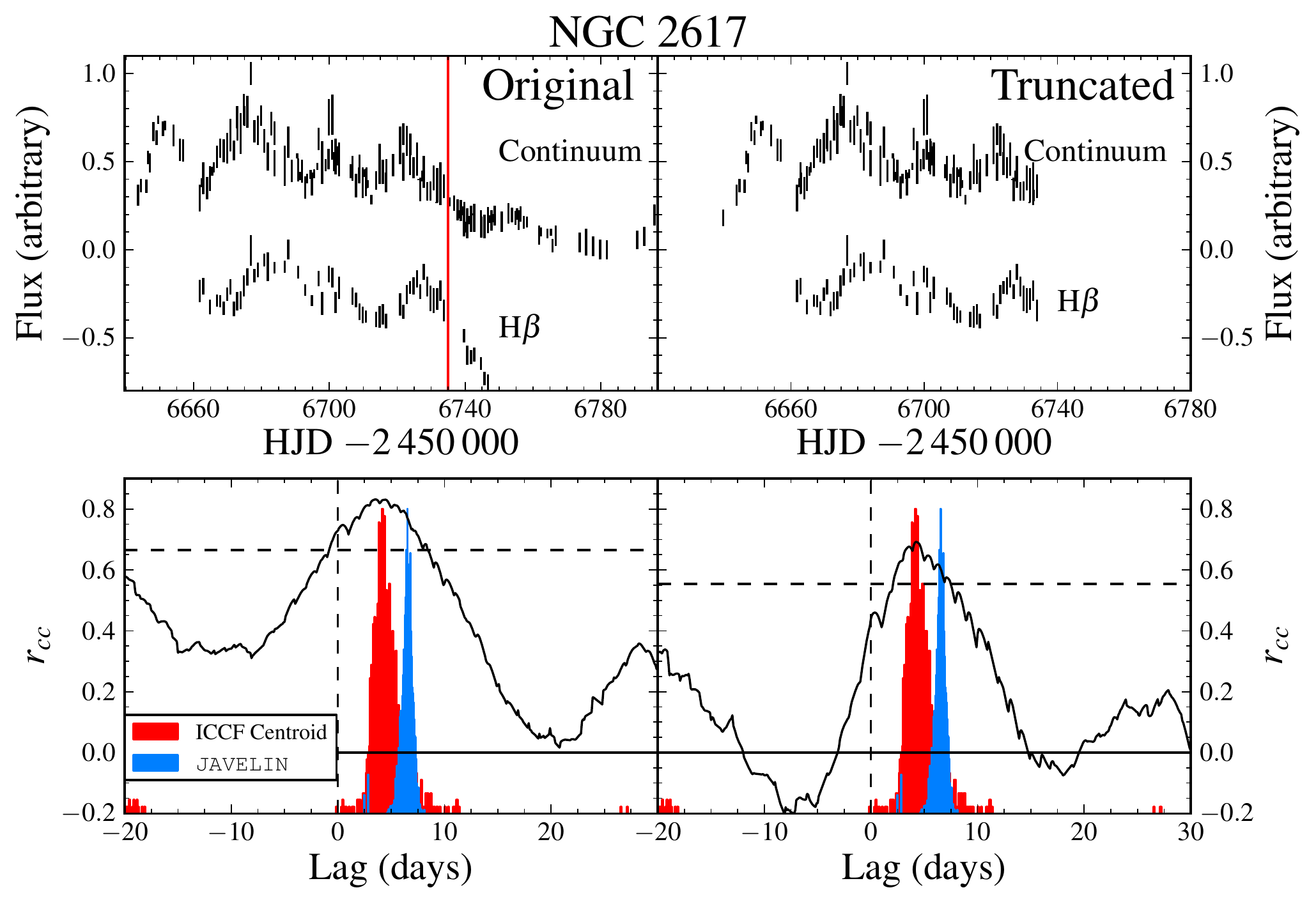}
\includegraphics[width=0.9\textwidth]{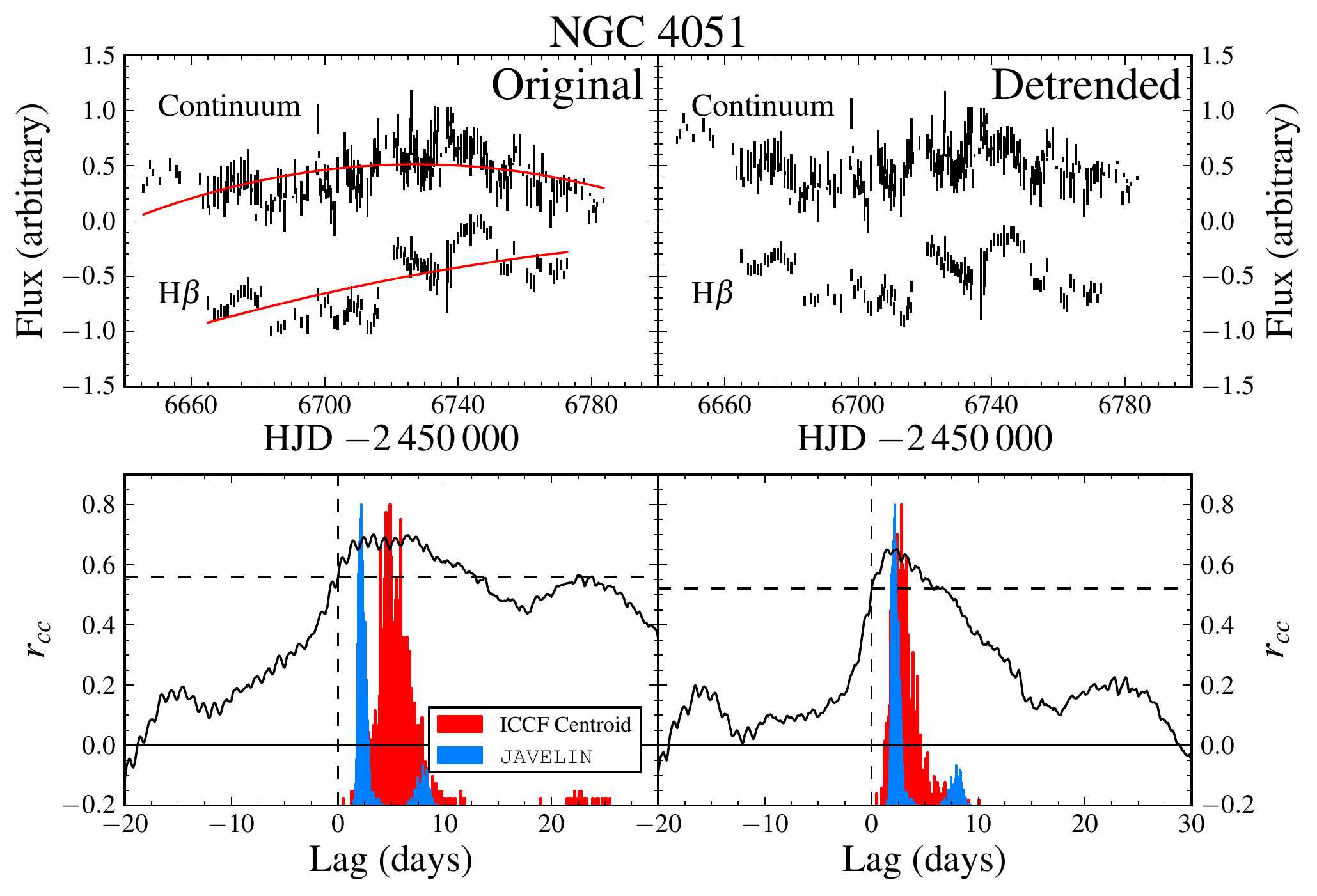}
\caption{Comparison of two CCF analyses of the H$\beta$ light curves
  for NGC\,2617 and NGC\,4051.  NGC\,2617 is shown in the top four
  panels and NGC\,4051 is shown in the bottom four.  Color coding of
  the histograms is the same as Figure\,\ref{fig:linelags}.  The left
  columns show the original light curves, the right columns show the
  modified light curves.  In the case of NGC\,2617, truncating the
  light curves at HJD$-2\,450\,000 = 6\,735$ days (vertical red line,
  top left) helps concentrate the ICCF peak.  For NGC\,4051,
  subtracting a second-order polynomial fit to the data removes the
  effect of long-term secular trends (red lines, bottom-middle left
  panel).  As a reminder, the {\tt JAVELIN} fits include linear
  detrending, so the blue histograms are identical on the left and
  right sides.  The same procedures were applied to all emission line
  light curves in these objects.\label{fig:detrend}}
\end{figure*}

We ran {\tt JAVELIN} models for each line using the 5100\,\AA\
continuum as the driving light curve, and we used internal {\tt
  JAVELIN} routines to remove any linear trends from the light curves
during the fit.  The damping time scale (a parameter of the DRW model)
for most AGN is several hundred days or longer \citep{Kelly2009,
  Macleod2010}, and our light curves are not long enough to
meaningfully constrain this parameter.  We therefore (arbitrarily)
fixed the damping time scale to 200 days.  We also tested several
different damping time scales (from a few days to 500 days), and found
that the choice of 200 days does not affect the best-fit lags---an
exact estimate of the damping time scale is not necessary to
reasonably interpolate the light curves \citep{Kozlowski2016b}.
Table\,\ref{tab:linelags} gives the median and 68\% confidence
interval of the posterior lag distributions, denoted as $\tau_{\tt
  JAV}$.  We also employed models that fit all light curves from a
single object simultaneously, which maximizes the available
information.  These results are given in Table\,\ref{tab:linelags} as
$\tau_{\rm multi}$.  Posterior distributions of $\tau_{\rm multi}$ are
shown by the blue histograms in Figure\,\ref{fig:linelags}.  For the
H$\gamma$ and \heii\ light curves from 3C\,382, we were again unable
to constrain any lag signal, and we drop these light curves from the
rest of this analysis.

\subsection{Results}

We generally find consistent results between the ICCF method and {\tt
  JAVELIN} models. The largest discrepancies are the H$\beta$ lags for
NGC\,2617 ($\Delta \tau = 1.6\sigma$) and 3C\,382 ($\Delta \tau
=2.0\sigma$), but these differences are not statistically significant.
In NGC\,2617, where the ICCF method detects a lag consistent with zero
in the H$\gamma$ or \heii\ light curves, {\tt JAVELIN} finds a lag at
reasonably high confidence: the percentiles for $\tau_{\rm multi}=0$
in the posterior lag distributions of H$\gamma$ and \heii\ are 8.3\%
and 1.1\%, which are 1.4$\sigma$ and 2.3$\sigma$ detections for
Gaussian probability distributions, respectively.  For Mrk\,374, an
H$\gamma$ lag is detected at high significance using {\tt JAVELIN} (we
do not claim a lag detection for \heii\ in this object, since the
$\tau_{\rm multi} = 0$ percentile is 20\%, only $0.2\sigma$ for a
Gaussian probability distribution).  The detection of these lags
represents a significant advantage of the {\tt JAVELIN} technique over
traditional cross-correlation methods.  We adopt the $\tau_{\rm
  multi}$ as our final lag measurements, since the multi-line global
fits provide well-constrained lags, properly treat covariances between
the lags from different light curves, and utilize the maximum amount
of information available in the data.

The analysis of NGC\,4051 is especially difficult because the light
curves exhibit low-amplitude variations. The lags in this object are
also expected to be small, based on the AGN luminosity
\citep{Bentz2013} and a previous well-sampled RM experiment
\citep{Denney2009b}.  For H$\beta$, {\tt JAVELIN} finds a definite lag
near 2 days, consistent with the detrended ICCF approach.  For
H$\gamma$, the ICCF method finds a lag consistent with zero, while the
single-line {\tt JAVELIN} fit finds a lag of $4.87 \pm 0.18$ days and
the multi-line fit finds a lag of $2.40\pm 0.80$ days (rest frame).
The single-line fit results in a complicated multi-modal posterior
distribution with smaller peaks at 15 and 25 days that are caused by
aliasing.  For example, the 25-day lag is probably caused by aligning
the H$\gamma$ maximum near 6745 days with the local maximum in the
continuum light curve at 6720 days (Figure\,\ref{fig:n4051}).
However, the multi-line fit shows a strong, dominant peak for
H$\gamma$ at $2.40$ days (rest frame). A probable explanation is that
the H$\beta$ light curve matches the overall shape of H$\gamma$, but
has stronger features against which to estimate a continuum
lag---fitting both light curves simultaneously can therefore establish
an H$\gamma$ lag with higher confidence.  The problem with the
H$\gamma$ light curve appears in a more serious form in the \heii\
light curve, and {\tt JAVELIN} finds a lag consistent with zero for
this line.

\section{Linewidths and $M_{\rm BH}$ Calculations}
After determining the characteristic size of the BLR from the mean
time delay, the next step is to calculate the characteristic
line-of-sight velocity of the BLR gas, from which we can derive SMBH
masses.  The BLR velocity is estimated from the width of emission
lines in the MDM spectra.  However, it is important to use the
linewidth of the variable component of the profile, since we measure
the BLR radius from the variable line flux.  For example, the variable
profile of 3C\,382 is radically different (and much broader) than the
time-averaged profile in the mean spectrum (Figure\,\ref{fig:3c382}).
We therefore measure and report in Table 16 linewidths both in the
mean spectrum $\hat F(\lambda)$, and in the rms spectrum $\sigma_{\rm
  var}(\lambda)$, but we use the latter for mass determinations.

There are two common choices for linewidth measurements: the
full-width at half-maximum (FWHM) and the line dispersion $\sigma_L$
(the rms width of the line profile).  There are advantages and
disadvantages associated with both approaches---while the FWHM is
simpler to measure, there are ambiguities for noisy or complicated
line profiles such as the double-peaked H$\beta$ profiles in
MCG+08-11-011, NGC\,2617, and 3C\,382.  On the other hand, although
$\sigma_L$ is well-defined for arbitrary line profiles, it depends
more sensitively on continuum subtraction and blending in the line
wings \citep{Denney2016, Mejia-Restrepo2016}.  \citet{Peterson2004}
find that velocities estimated with $\sigma_L$ produce a tighter
virial relation, and \citet{Denney2013} find that the masses
determined from UV and optical lines agree better using $\sigma_{\rm
  L}$.  We therefore adopt $\sigma_L$ as a measure of the BLR velocity
in this study.  For completeness, we also give the FWHM in
Table\,\ref{tab:v_alt}.

Linewidth uncertainties are estimated using a bootstrapping method.
For $10^3$ iterations on each object with $N$ nightly spectra, we
randomly select $N$ observations with replacement, recompute the mean
and rms spectrum, and remeasure the linewidths in the rms spectrum.
The central 68\% confidence interval of the resulting distributions
are adopted as the formal uncertainty of the linewidth.  This approach
can only account for statistical uncertainties in the linewidths,
which therefore represent lower limits on the uncertainties.  There
are additional systematic errors from the choice of wavelength windows
that define the line profiles (Tables \ref{tab:windows} and
\ref{tab:con_windows}), as well as blending of the broad-line wings.
The choice of wavelength windows and continuum subtraction is
problematic for weak lines, lines with low variability, and lines with
unusual profiles.  In particular, our estimates for the \heii\ line in
NGC\,2617, NGC\,4051, 3C\,382, and all lines in Mrk\,374 are certainly
affected.  Furthermore, the blue wing of H$\beta$ and the red wing of
\heii\ overlap in MCG+08-11-011 and NGC\,2617, and it is likely that
the \heii\ velocity is severely underestimated (the effect on H$\beta$
is probably smaller, though it may not be negligible).  Spectral
decompositions may help with these problems in future analyses; for
now, we note that the linewidth uncertainties are underestimated in
these cases, and we provide
 a treatment for this issue below.

We correct the linewidth measurements for the instrument resolution by
subtracting the rms width of the spectrograph's line-spread-function
(LSF) in quadrature from the observed value of $\sigma_L$.  Previous
studies have found that the width of the LSF for the MDM spectrograph
is near 3.2 or 3.4\,\AA\ (FWHM 7.6--7.9\,\AA,
\citealt{Denney2010,Grier2012}).  Based on comparisons with high
spectral resolution observations, where the LSF width is negligible,
we find a LSF width of 2.97\,\AA\ (FWHM $=6.99\,\AA$).  This value was
determined using the catalog of high-resolution \oiii\ measurements
from \citet{Whittle1992}, which contains intrinsic \oiii\ linewidths
for MCG+0-11-011 and NGC\,4051.  The \oiii\ line of NGC 4051 is
undersampled in the MDM spectra (the intrinsic FWHM is 190 km
s$^{-1}$, or 3.16\,\AA\ in the observed frame), and does not give a
reliable estimate the instrumental broadening.  However, the intrinsic
\oiii\ FWHM in MCG+08-11-011 is 605 km s$^{-1}$, or 10.52\,\AA\ in the
observed frame, which is well resolved.  The observed FWHM in the
MCG+08-11-011 reference spectrum (before smoothing, see \S2.2.2 and
below) is 12.63\,\AA, which implies that the FWHM of the LSF is
6.99\,\AA\ (a rms width of 2.97\,\AA).  This value is close to but
slightly smaller than previous estimates.  The MDM LSF may not be
perfectly stable in time, so we adopt 2.97\,\AA\ as the rms width of
the instrumental broadening in our observations.

An additional correction must be applied because we smooth our
reference spectra to approximately match the nights with the worst
spectroscopic resolution (see \S2.2.2).  The kernel widths for this
smoothing procedure were 1.4\,\AA\ for MCG+08-11-011, 1.5\,\AA\ for
NGC\,2617, 1.8\,\AA\ for NGC\,4051, 1.7\,\AA\ for 3C\,382, and
1.9\,\AA\ for Mrk\,374 (the FWHM values are a factor of 2.35 larger).
We also subtract these values in quadrature from the observed line
dispersion.  The final rest-frame linewidths and their uncertainties
are given in Table\,\ref{tab:v_alt}.

\floattable
\begin{deluxetable}{llrrrr}
\tablewidth{0pt}
\tablecaption{Rest-frame Velocity Linewidth Measurements \label{tab:v_alt}}
\tablehead{
& &\multicolumn{2}{c}{RMS Spectrum} & \multicolumn{2}{c}{Mean Spectrum}\\
\colhead{Object} & \colhead{Line} & \colhead{$\sigma_{\rm L}$} & \colhead{FWHM} & \colhead{$\sigma_{\rm L}$} & \colhead{FWHM} \\
& &\colhead{(km s$^{-1}$)}&\colhead{(km s$^{-1}$)}&\colhead{(km s$^{-1}$)}&\colhead{(km s$^{-1}$)}\\
\colhead{(1)}&\colhead{(2)}&\colhead{(3)}&\colhead{(4)}&\colhead{(5)}&\colhead{(6)}
}
\startdata
MCG+08-11-011 & H$\beta$ & $ 1466_{-174}^{+102}$ & $ 4475_{-356}^{+192}$ & $ 1681_{-02}^{+02}$ & $ 1159_{-0007}^{+0008}$ \\
 & H$\gamma$ & $ 1604_{-082}^{+083}$ & $ 3916_{-716}^{+616}$ & $ 1175_{-07}^{+07}$ & $ 1978_{-0019}^{+0026}$ \\
 & He{\sc ii}$\lambda$4686 & $ 2453_{-130}^{+125}$ & $ 6617_{-776}^{+993}$ & $ 2893_{-37}^{+42}$ & $ 2517_{-1020}^{+1814}$ \\
\hline
NGC 2617 & H$\beta$ & $ 2424_{-086}^{+091}$ & $ 6489_{-162}^{+213}$ & $ 2709_{-006}^{+006}$ & $ 5303_{-46}^{+49}$ \\
 & H$\gamma$ & $ 3084_{-090}^{+086}$ & $ 7674_{-472}^{+423}$ & $ 2385_{-034}^{+033}$ & $ 4101_{-29}^{+26}$ \\
 & He{\sc ii}$\lambda$4686 & $ 2020_{-572}^{+329}$ & $ 6788_{-855}^{+984}$ & $ 3113_{-218}^{+147}$ & $ 7150_{-33}^{+35}$ \\
\hline
NGC 4051 & H$\beta$ & $  493_{-36}^{+34}$ & $  941_{-019}^{+017}$ & $  470_{-2}^{+2}$ & $  765_{-03}^{+03}$ \\
 & H$\gamma$ & $  641_{-59}^{+55}$ & $ 1098_{-034}^{+031}$ & $  942_{-4}^{+4}$ & $ 1676_{-05}^{+06}$ \\
 & He{\sc ii}$\lambda$4686 & $ 1689_{-38}^{+36}$ & $ 3885_{-218}^{+299}$ & $ 1898_{-3}^{+4}$ & $ 4598_{-11}^{+10}$ \\
\hline
3C382 & H$\beta$ & $ 4552_{-0163}^{+0214}$ & $11549_{-0667}^{+1292}$ & $ 3227_{-07}^{+07}$ & $ 3619_{-050}^{+282}$ \\
 & H$\gamma$ & $ 5083_{-1942}^{+1114}$ & $10706_{-1294}^{+1050}$ & $ 2845_{-53}^{+55}$ & $ 3483_{-022}^{+022}$ \\
 & He{\sc ii}$\lambda$4686 & $ 2073_{-0352}^{+0170}$ & $ 4374_{-0602}^{+1257}$ & $ 1789_{-28}^{+27}$ & $ 5186_{-075}^{+247}$ \\
\hline
Mrk 374 & H$\beta$ & $ 1329_{-429}^{+308}$ & $ 3094_{-0619}^{+0488}$ & $ 1490_{-04}^{+04}$ & $ 3250_{-18}^{+19}$ \\
 & H$\gamma$ & $ 1163_{-364}^{+215}$ & $ 2311_{-0294}^{+0814}$ & $ 1148_{-05}^{+05}$ & $ 3648_{-18}^{+18}$ \\
 & He{\sc ii}$\lambda$4686 & $ 1997_{-223}^{+158}$ & $ 4172_{-1083}^{+1106}$ & $ 1554_{-70}^{+58}$ & $ 4140_{-64}^{+64}$ \\
\enddata
\tablecomments{Column 3 and Column 4 give the rms line width and FWHM
  in the rms spectrum.  Column 5 and Column 6 give the same but in the
  mean spectrum. All values are corrected for instrumental broadening
  and the smoothing in \S2.2.2 (see \S4), and are reported in the
  rest-frame.  The uncertainties only represent the statistical
  errors---blending, continuum interpolation, and the choice of
  wavelength windows introduce additional systematic uncertainties
  (especially for \heii).}
\end{deluxetable}

We measure the SMBH masses as
\begin{align}
  M_{\rm BH} = \langle f\rangle \frac{\sigma_L^2c\tau_{\rm multi}}{G}
\end{align}
where $c$ is the speed of light, $G$ is the gravitational constant,
and $\langle f\rangle $ is the virial factor.  The virial factor
accounts for the unknown geometry and dynamics of the BLR, and is
determined by calibrating a sample of RM AGN to the $M_{\rm
  BH}$-$\sigma_{*}$ relation (e.g., \citealt{Onken2004, Park2012b,
  Grier2013b}).  We use the most recent calibration by \citet{Woo2015}
of $\langle f\rangle = 4.47\pm 1.25$ with a scatter of $0.43\pm 0.03$\
dex (a factor of 2.7).  Finally, it is convenient to define the virial
product, $\sigma_L^2c\tau/G$, which is an observed quantity that is
independent of the mass calibration.

We calculate the statistical uncertainties on the virial
  products through standard error propagation.  As discussed above,
  there are significant systematic uncertainties on both the
  linewidths and the lags, which probably dominate the final error
  budget (see also \S2.4).  We estimate the systematic uncertainty
  using repeat RM measurements gathered from the literature.  There
  are 17 H$\beta$-based measurements of the virial product in
  NGC\,5548 over the last 30 years (see \citealt{Bentz2015}).  The
  (log) standard deviation of these measurements is 0.16 dex, while
  the mean statistical uncertainty is 0.10 dex.  Taking $\sigma_{\rm
    sys}^2 = \sigma_{\rm rms}^2 - \sigma_{\rm stat}^2$, we estimate a
  systematic uncertainty floor of 0.13 dex.  Experimentation with
  alternative line windows, continuum interpolations, and detrending
  procedures suggests that this value (a factor of about $1.3$)
  captures most of the variation in the virial products of our sample.
  We therefore adopt 0.13 dex as our estimate of the systematic
  uncertainty on each virial product, and add this value in quadrature
  to the statistical uncertainties for the virial products.  For our
  final mass estimates, we also add in quadrature the the uncertainty
  in the mean value of $\langle f\rangle $ ($\sim\!0.12$ dex) and its
  intrinsic scatter (0.43 dex).  The virial products, final masses,
  and total uncertainties are given in Table~\ref{tab:masses}.  

We discuss the consistency of virial products for the same
  object derived from different emission lines in \S5.2, and we
  comment on the H$\beta$-derived masses of individual objects below.

 \floattable
\begin{deluxetable}{llrrrrcrr}
\tablewidth{0pt}
\tablecaption{Black Hole Masses \label{tab:masses}}
\tablehead{\colhead{Object} & \colhead{Line} & \colhead{$\tau$}  & \colhead{$\sigma_{L}$}  & \colhead{$\log \left(c\tau \sigma^2_{V}/G\right) $\tablenotemark{a}} & \colhead{$\log M_{\rm BH} $\tablenotemark{b}}\\
& &\colhead{(days)}&\colhead{(km s$^{-1}$)}&\colhead{[M$_{\odot}$]}&\colhead{[M$_{\odot}$]}\\
\colhead{(1)}&\colhead{(2)}&\colhead{(3)}&\colhead{(4)}&\colhead{(5)}&\colhead{(6)}
}
\startdata
MCG+08-11-011 & H$\beta$ & $14.98^{+0.34}_{-0.28}$ & $ 1466^{+102}_{-174}$ & $6.80 \pm 0.15$ & $7.45 \pm 0.47$ \\
 & H$\gamma$ & $12.38^{+0.46}_{-0.49}$ & $ 1604^{+ 83}_{- 82}$ & $6.79 \pm 0.14$ & $7.44 \pm 0.47$ \\
 & He{\sc ii}$\lambda$4686 & $1.21^{+0.29}_{-0.33}$ & $ 2453^{+125}_{-130}$ & $6.20 \pm 0.18$ & $6.80 \pm 0.48$ \\
\hline
NGC 2617 & H$\beta$ & $6.38^{+0.44}_{-0.50}$ & $ 2424^{+ 91}_{- 86}$ & $6.86 \pm 0.14$ & $7.51 \pm 0.47$ \\
 & H$\gamma$ & $0.81^{+0.59}_{-0.61}$ & $ 3084^{+ 86}_{- 90}$ & $6.17 \pm 0.44$ & $6.82 \pm 0.63$ \\
 & He{\sc ii}$\lambda$4686 & $1.75^{+0.34}_{-0.38}$ & $ 2020^{+329}_{-572}$ & $6.14 \pm 0.26$ & $6.79 \pm 0.52$ \\
\hline
NGC 4051 & H$\beta$ & $2.24^{+0.39}_{-0.28}$ & $  493^{+ 34}_{- 36}$ & $5.02 \pm 0.16$ & $5.67 \pm 0.47$ \\
 & H$\gamma$ & $2.40^{+0.86}_{-0.73}$ & $  641^{+ 55}_{- 59}$ & $5.28 \pm 0.21$ & $5.93 \pm 0.50$ \\
\hline
3C382 & H$\beta$ &  $52.07^{+3.18} _{-9.46}$ & $ 4552^{+214}_{-163}$ & $8.33 \pm 0.14$ & $8.98 \pm 0.47$ \\
\hline
Mrk 374 & H$\beta$ & $13.73^{+1.06}_{-1.02}$ & $ 1329^{+308}_{-429}$ & $6.67 \pm 0.31$ & $7.32 \pm 0.54$ \\
& H$\gamma$ & $13.37^{+2.11}_{-2.08}$ & $ 1163^{+215}_{-364}$ & $6.55 \pm 0.28$ & $7.20 \pm 0.53$ \\
\enddata
\tablenotetext{a}{Includes a 0.13 dex systematic uncertainty, added in
  quadrature to the statistical uncertainties propagated from Columns
  3 and 4.  }  
\tablenotetext{b}{Include uncertainty in the mean value
  of $f$ (0.12 dex) and its intrinsic scatter (0.43 dex) added in
  quadrature to the uncertainties from Column 5.}
\tablecomments{Column 3 gives the adopted lag and its statistical
  uncertainty, $\tau_{\rm multi}$, from Table \ref{tab:linelags}.
  Column 4 gives the rms linewidth $\sigma_{\rm L}$ from Table
  \ref{tab:v_alt} of the line profile in the rms residual spectrum and
  its statistical uncertainty (see \S2.4 and \S4), corrected to the
  rest-frame.  Column 5 gives the virial product, which is independent
  of any calibration to the $M_{\rm BH}$--$\sigma_{*}$ relation.
  Column 6 gives the SMBH mass using the $M_{\rm BH}$--$\sigma_{*}$
  calibration from \citet{Woo2015} with$f =4.47\pm 1.25 $. }
\end{deluxetable}

\begin{enumerate}[i.]
\item MCG+08-11-011 is our most variable object.  The black hole mass
  estimate is $\sim\!2.8 \times 10^7\, {\rm M_\odot}$, and the
  uncertainty is dominated by uncertainty in the virial factor $f$.
  \citet{Bianchi2010} found evidence for a relativistically broadened
  Fe K$\alpha$ line in the X-ray spectrum of this object, but the
  available mass estimates at that time were uncertain by an order of
  magnitude (10$^7$--10$^8$\,M$_\odot$).  The results presented here
  may help measure the spin of the black hole in future studies.

\item The mass reported here for NGC\,2617 of $\sim\!3.2 \times
  10^{7}\, {\rm M_{\odot}}$ is in good agreement with the single-epoch
  mass estimated by \citet{Shappee2014} of $(4 \pm 1) \times 10^{7}\,
  {\rm M_{\odot}}$, also using the H$\beta$ emission-line.  NGC\,2617
  is the second ``changing look'' AGN with a direct RM mass
  measurement.  The other object is Mrk 590, which was observed to
  change from a Seyfert 1.5 to 1.0 to 1.9 over several decades
  \citep{Denney2014}, and has a RM mass of $\sim\!  5\times
  10^{7}\,{\rm M_{\odot}}$ \citep{Peterson2004}.  In terms of their
  black hole masses, there is nothing extraordinary about either NGC
  2617 or Mrk 590.  Our luminosity-independent RM mass also allows us
  to estimate a more robust Eddington ratio ($\dot m_{\rm Edd} =
  L_{\rm Bol}/L_{\rm Edd}$) than from the single-epoch mass.  Assuming
  a bolometric correction of 10 for the 5100\,\AA\ continuum
  luminosity, we find that $\dot m_{\rm Edd} = 0.01$, after correcting
  for host-galaxy starlight (see \S5.1).  This value is somewhat low,
  though not atypical, for Seyfert 1 galaxies.

\item For NGC\,4051, our measurement of the H$\beta$ lag ($2.24 \pm
  0.33$ days) is in good agreement with the estimate of $1.87\pm 0.52$
  days by \citet{Denney2009b}.  The measurement is challenging because
  of the low-amplitude continuum variations, variable host-galaxy
  contamination from aperture effects \citep{Peterson1995}, and a
  secular trend in the line light curve.

  Our estimate of the virial product $\sim\! 1.1 \times 10^5
  $\,M$_\odot$\ is also consistent at the 2$\sigma$ level with the
  estimate of $(3.0\pm 1.0) \times 10^5$\,M$_\odot$\ from
  \citet{Denney2010}.  The difference is primarily due to a decrease
  in the linewidth by about 400 ${\rm km\ s^{-1}}$ compared to the
  2007 campaign.  The line and continuum wavelength window definitions
  are somewhat different between the 2014 and 2007 campaigns, and we
  found that using the wavelength windows from Tables
  \ref{tab:windows} and \ref{tab:con_windows} for the rms spectrum
  from 2007 reduces the difference to only $\sim\!100 {\rm \ km\
    s^{-1}}$ (i.e., $\sigma_{\rm L}$ was about 20\% larger in 2007
  than in 2014).  If we use the wavelength regions from
  \citet{Denney2010}, the measurement from 2014 increases by $\sim\!
  120 {\rm\ km\ s^{-1}}$.  This suggests that the virial product is
  somewhat smaller than that reported by \citet{Denney2010}, but the
  mild 2$\sigma$ discrepancy indicates that the systematic
  uncertainties are comparable to the formal uncertainties.  The
  remaining 100--300 ${\rm\ km\ s^{-1}}$\ difference is
  physical---comparing the rms line profiles between the two
  campaigns, we found that the core of the H$\beta$ line is much more
  variable in 2014 than it was in 2007, weighting $\sigma_{\rm L}$ to
  smaller values.  The lag has only increased by 0.26 days (19\%), so
  the virial product shows a net decrease.  This might indicate a
  change in the geometry and/or dynamics of the BLR.  The dynamical
  time is of order only two or three years at two light days from a
  $10^6$ M$_\odot$ black hole, so such a change cannot be ruled out
  {\it a priori}.  A comparison of the velocity resolved reverberation
  signals between 2007 and 2014 is therefore especially interesting.

  Our SMBH mass estimate of $\sim \! 4.7 \times 10^5$ ${\rm M_{\odot}}$
  for NGC\,4051 is at the very low end of the SMBH scale, and there
  are only two other RM masses below $10^{6}$ $\rm{M_{\odot}}$: NGC
  4395 \citep{Peterson2005,Edri2012} and UGC 06728 \citep{Bentz2016}.

\item In 3C\,382 the black hole mass is about $9.6 \times 10^8$ ${\rm
    M_{\odot}}$, and a large source of uncertainty is the H$\beta$
  lag.  The $\sim$\,52 day lag is driven by the gentle inflection in
  the line light curve observed near the middle of the spectroscopic
  campaign, which was also observed in the imaging data about one
  month before the MDM observations began.  The uncertainties on the
  H$\beta$ line lag are therefore quite large.  By RM standards,
  3C\,382 is also at a moderate redshift ($z \sim 0.06$) and faint
  ({\it V} $\sim\!  15.4 $), putting it near the limit of feasibility
  for monitoring campaigns with a 1m-class telescope.

  Several estimates of the BLR orientation exist for this
  object. Emission from the radio lobes in 3C\,382 dominates over that
  of the core, indicating that the system is viewed more edge on
  (\citealt{Wills1986} give the core-to-lobe ratio as $\sim\! 0.1$).
  However, \citet{Eracleous1995} find an inclination of 45$^{\circ}$
  from dynamical modeling of the double-peaked broad H$\alpha$ line
  and show that this estimate is consistent with the radio properties.
  Velocity-delay maps and dynamical modeling of this object would be
  an interesting test of this inclination measurement.  Unfortunately,
  the width of the continuum autocorrelation function and the low S/N
  of the line light curves are poorly suited for these experiments.
   On the other hand, a moderately inclined disk is broadly
    consistent with the double-peaked rms H$\beta$ and H$\gamma$ line
    profiles, and velocity-binned mean time delays may still provide
    interesting constraints on the BLR structure.

\item Mrk\,374 is our least variable source.  Although the H$\beta$
  lag is detected at a statistically significant level, the
  uncertainty on the ICCF centroid is somewhat larger than for the
  other objects ($\sim\!  33\%$).  The mass is $\sim\!2.09\times 10^7$
  ${\rm M_{\odot}}$, and the dominant uncertainty is from the
  linewidth measurement---it is clear from Figure \ref{fig:mrk374}
  that the variability of the lines is very small and that there is
  some ambiguity in where the line profile begins and ends. At a
  redshift of $\sim\!  0.04$, Mrk\,374 is one of our fainter sources
  ({\it V}$ = 15.0$ mag), and, similar to 3C\,382, it is near the
  practical limits of a monitoring campaign lead by a 1m-class
  telescope.
\end{enumerate}

\section{Discussion}
\subsection{Radius-Luminosity Relation}
\begin{figure}
\includegraphics[width=0.5\textwidth]{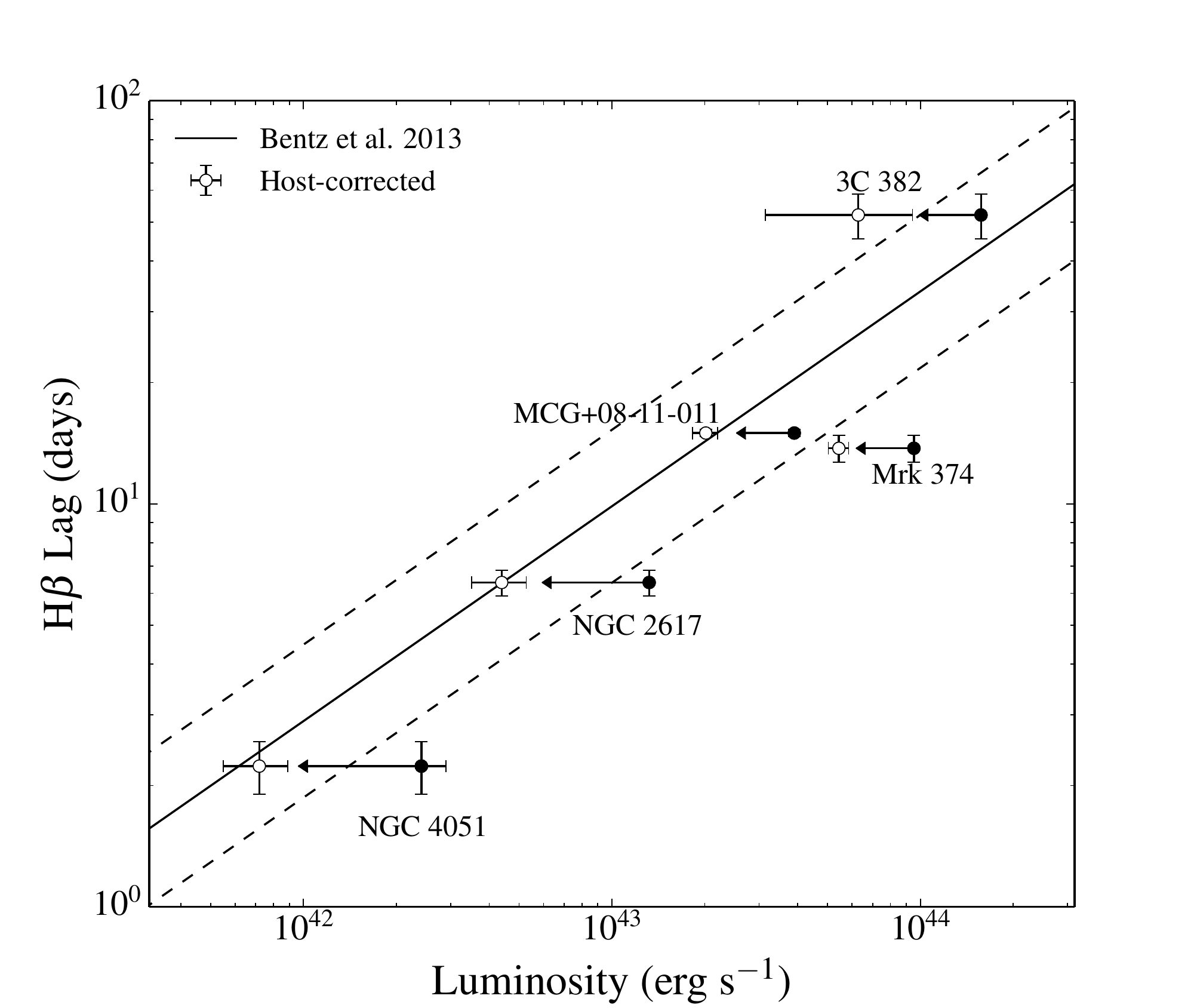}
\caption{Radius--luminosity relation for the targets of this study,
  compared to the relation from \citet{Bentz2013}.  Luminosities are
  estimated from the mean of the continuum light curves corrected for
  Galactic extinction.  The solid black line shows the best-fit
  relation measured by \citet{Bentz2013}, and the dashed black lines
  show the dispersion around the best fit.  Open circles show the
  luminosities corrected for host-galaxy starlight, which results in
  excellent agreement with the relation from
  \citet{Bentz2013}.  \label{fig:RL}}
\end{figure}

Figure\,\ref{fig:RL} shows the H$\beta$ lags of our five objects as a
function of luminosity, the so-called radius-luminosity ($R$--$L$)
relation (\citealt{Kaspi2000,Kaspi2005, Bentz2009,Bentz2013}).  To
estimate the luminosities, we first take the mean of the 5100\,\AA\
light curve and correct for Galactic extinction using the extinction
map of \citet{Schlafly2011} and a \citet*{Cardelli1989} extinction law
with $R_V = 3.1$.  We then convert the flux to luminosity using the
luminosity distances in Table\,\ref{tab:targets}.  In the case of
NGC\,4051, which has a large peculiar velocity relative to the Hubble
flow ($z \sim 0.002$), we use a Tully-Fischer distance of 17.1 Mpc
\citep{Tully2008}.  This distance is uncertain by about 20\%, and
improving this measurement is an important step to investigate any
discrepancies of this object from the $R$--$L$ relation and to
estimate its true Eddington ratio.  For these purposes, an {\it HST}
program has recently been approved to obtain a Cepheid distance to
NGC\,4051 ({\it HST} GO-14697; PI Peterson).

The final values of $\lambda L_{5100\angstrom}$ are reported in Table
\ref{tab:targets}, along with the adopted Galactic values of $E(B-V)$.
We find that our objects all lie close to, but slightly below (except
for 3C\,382), the $R$--$L$ relation.  The major systematic
uncertainties are internal extinction in the AGN and host-galaxy
contamination.  Internal extinction may move the points farther from
the $R$--$L$ relation, but this effect is expected to be small.  On
the other hand, host-galaxy contamination can be very significant,
especially for low-luminosity objects.

In order to correct for host contamination, we model high-resolution
images of the targets and isolate the host-galaxy flux.  This has
previously been done for NGC\,4051 \citep{Bentz2006, Bentz2013}, and
MCG+08-11-011, NGC\,2617, and Mrk\,374 were recently observed with
{\it HST} for this purpose ({\it HST} GO-13816; PI Bentz).  We also
retrieved archival optical WFPC2 imaging of 3C\,382 ({\it HST}
GO-6967, PI Sparks), but the data are not ideal for image
decompositions and we discuss the host-galaxy flux estimate for this
object separately.  A more detailed analysis of the {\it HST} GO-13816
data and image decompositions will be presented in future work (Bentz
et al, in preparation).  However, following the procedures described
by \citet{Bentz2013}, we made preliminary estimates of the host-galaxy
contributions in the MDM aperture ($15\farcs 0 \times 5\farcs 0$
aligned at position angle $0^{\circ}$).  The results are given in
Table\,\ref{tab:targets} (uncertainties on these values are estimated
at 10\% and included in Figure\,\ref{fig:RL}).  Applying this
correction shows that host-contamination accounts for the entire
discrepancy between the observed luminosities and the $R$--$L$
relation.  The largest deviation from the $R$--$L$ relation is Mrk\,374,
but the offset is only slightly greater than the $1\sigma$ scatter of
the relation.

3C\,382 resides in a giant elliptical galaxy and there may be a
significant contribution from the host's starlight---several stellar
absorption features are visible in the mean spectrum in Figure
\ref{fig:3c382}.  In the archival {\it HST} images, the galaxy nucleus
is saturated, hindering our ability to robustly remove the AGN flux
and isolate the host's starlight.  The main problem is that the Sersic
index of the host-galaxy is degenerate with the saturated core and
tends to drift toward unreasonably high values ($n\approx 7.6$) when
fitting the image in the same way as \citet{Bentz2013}.  Fixing the
Sersic index to more typical values (between 2 and 4) leads to host
fluxes in the MDM aperture between 2.2 and 2.7$\times 10^{-15}$ erg
cm$^{-2}$ s$^{-1}$ \AA$^{-1}$, about 77\% of the observed luminosity
($\log \lambda L_{\rm host} = 44.04$ to 44.12 [erg s$^{-1}$], after
correcting for Galactic extinction).  This estimate can be checked
using the equivalent-width (EW) of the prominent Mg absorption feature
at 5200\,\AA\ rest-frame (5460\,\AA\ observed-frame).  In our mean
spectrum, we find an EW of 2.8\,\AA.  In typical elliptical galaxy
spectra, we find the EW is about 6.7 to 7.3\,\AA, depending on the
continuum estimation and assumptions about the host-galaxy
properties.\footnote{We used two different templates for the
  ``standard'' giant elliptical spectrum: observations of the E0
  galaxy NGC 1407 used to construct empirical templates
  \citep{Kinney1996,Denney2009a}, and a synthetic stellar population
  model from \citet{Bruzual2003} consisting of a single 11 Gyr
  population at solar metallicity.} This implies that the featureless
AGN continuum dilutes the absorption feature by a factor of 2.4 to
2.6, so that the host galaxy contributes approximately 40\% of the
observed luminosity.  This rough estimate is a factor of two lower
than the result from image decomposition, but the two values span the
range of host-contributions from the other objects in our sample (42\%
to 71\% of the observed luminosity, see Table \ref{tab:targets}).  We
therefore adopt a host correction of $(60 \pm 20)$\% of the observed
luminosity ($\log \lambda L_{\rm 5100\angstrom} = 43.98 \pm 0.15$ [erg
s$^{-1}$]), and we note that this estimate can easily be improved by
obtaining unsaturated high resolution images.  The host correction
moves 3C\,382 away from the $R$--$L$ relation, just beyond the
$1\sigma$ dispersion.  However, considering the large uncertainties,
there does not appear to be any evidence that 3C\,382 has an anomalous
H$\beta$ lag for its luminosity.

\subsection{Virialization of the BLR}
\begin{figure*}
\includegraphics[width=\textwidth]{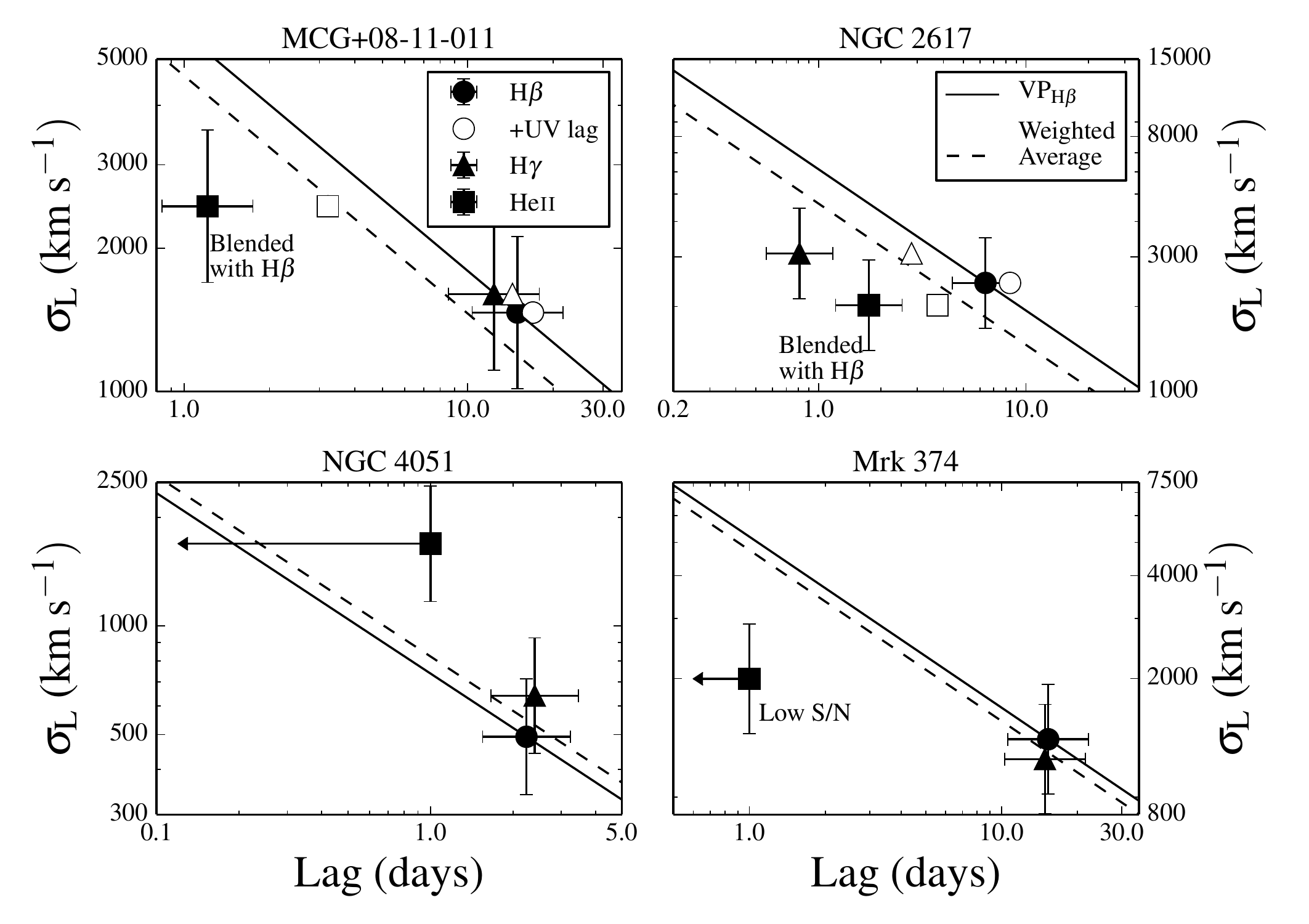}
\caption{Linewidth $\sigma_{\rm L}$ as a function of lag $\tau_{\rm
    multi}$ from Columns 3 and 4 of Table\,\ref{tab:masses}.  The
  solid lines show the virial relation $\sigma_{\rm L}\propto
  \tau_{\rm multi}^{-1/2}$ normalized by the H$\beta$ virial product
  (VP$_{\rm H\beta}$).  The dashed lines show the virial relation
  using a weighted average of the different emission line virial
  products (Column 5 of Table \ref{tab:masses}).  The open points for
  MCG+08-11-011 and NGC\,2617 show the effect of adding a hypothetical
  2.0 day UV--optical lag, similar to that found in NGC\,5548 (see
  \S5.2).  For NGC\,4051 and Mrk\,374, the \heii\ lags are consistent
  with zero, and we show these as upper limits at 1 day.
  Uncertainties of 0.13 dex are assigned to both $\tau_{\rm multi}$
  and $\sigma_{\rm L}$, to approximately represent the level of
  systematic uncertainty associated with the virial
  products.  \label{fig:virial}}
\end{figure*}

With the measurement of BLR velocity dispersions at a range of radii,
it is possible to test if the BLR is virialized.  Virialized dynamics
predict $V(r) \propto r^{-1/2}$, where the constant of proportionality
depends on the SMBH mass and BLR inclination/kinematics.  If the BLR
is virialized, the virial products $\sigma_L^2c\tau/G$ derived from
different line species should be consistent with each other, assuming
similar geometries and dynamics for the line-emitting gas.

In Table\,\ref{tab:masses}, the maximum differences between
  $\log \sigma_L^2c\tau/G$ for each object are $3.3\sigma$ in
  MCG+08-11-011, $2.8\sigma$ in NGC\,2617, $1.2\sigma$ in NGC\,4051,
  and $0.4\sigma$ in Mrk\,374.  For NGC\,4051 and Mrk\,374, these
differences are not significant.  For MCG+08-11-011 and NGC 2617, the
H$\beta$ and \heii\ virial products are marginally discrepant at about
2.5--3.3$\sigma$. We show these results Figure\,\ref{fig:virial},
which displays the linewidths $\sigma_{\rm L}$ as a function of lag
$\tau_{\rm multi}$, and the relation $\sigma_{\rm L}\propto \tau_{\rm
  multi}^{-1/2}$ normalized by the value for H$\beta$.  In
  this figure, we have applied a 0.13 dex uncertainty to both the lag
  $\tau$ and line width $\sigma_{\rm L}$, representative of the
  characteristic systematic uncertainties.  While the H$\gamma$
points generally agree with the H$\beta$ relation, the \heii\ points
have very large offsets.

There are many systematic issues that could account for these
differences.  As discussed in \S4, the red wing of \heii\ is blended
with the blue wing of H$\beta$ in both MCG+08-11-011 and NGC\,2617.
The \heii\ velocity is therefore likely underestimated because we
cannot follow its red wing underneath H$\beta$.  The \heii\ lags are
also small compared to the monitoring cadence, and the lag is only
marginally detected at 2.3$\sigma$ in NGC\,2617.  Furthermore,
  the choice of line window and continuum interpolation can have a
  significant effect on the lag and linewidths.  Finally, we must
assume that the 5100\,\AA\ continuum light curve is a suitable proxy
for the ionizing flux variations at extreme UV wavelengths.  In
NGC\,5548, we found a $\sim\!  2$ day lag between the far UV and
optical emission \citep{Edelson2015,Fausnaugh2016}.  If a similar lag
exists in these objects, it would change the \heii\ virial products by
a significant amount (0.3--0.4 dex), while the change in the H$\beta$
virial products would be much smaller (0.05--0.11 dex).  The effect of
adding a 2 day UV-optical lag to the optical-line lags is shown in
Figure \ref{fig:virial}, and the additional lag would reduce the
discrepancies in the virial products to 1.3$\sigma$ for MCG+08-11-011
and 2.0$\sigma$ for NGC\,2617.  These AGN have masses and luminosities
similar to NGC\,5548, so the existence of a UV-optical lag of this
magnitude is very likely.  Although a UV-optical lag affects the
virial product and the characteristic size of the BLR, it does not
affect the final mass estimate because the virial factor $\langle f
\rangle$ is calibrated using the $M_{\rm BH}$--$\sigma_{*}$ relation
(see \citealt{Fausnaugh2016,Pei2017}).

If the remaining discrepancies are real, they indicate different
dynamics and geometries for the \heii\ line-emitting gas compared to
that of H$\beta$.  This might be plausible, since \heii\ is a
high-ionization state line and may originate in very different
physical conditions than the Balmer lines (for example, a disk wind).
If \heii\ has different dynamics than H$\beta$, it would be necessary
to calibrate a different virial factor $\langle f \rangle$ for the
\heii\ line when calculating the SMBH masses.  However, we cannot rule
out systematic effects and it is unclear if the \heii\ discrepancies
are physical.  If systematic issues do account for the discrepancies,
then the dynamics of the BLRs in these AGN would be consistent with
virialized motion, as has been found for other AGN
\citep{Peterson2004}.

The H$\beta$ light curves and line profiles have much higher S/N and
very clear lags compared to both \heii\ and H$\gamma$, resulting in
more reliable black hole masses.  If we combine the virial products in
Table\,\ref{tab:masses} using an error-weighted average, the virial
relation changes little, as shown in Figure\,\ref{fig:virial} with the
dashed lines.  We therefore take the H$\beta$ masses for our standard
SMBH mass estimates.

\begin{figure*}
\includegraphics[width=\textwidth]{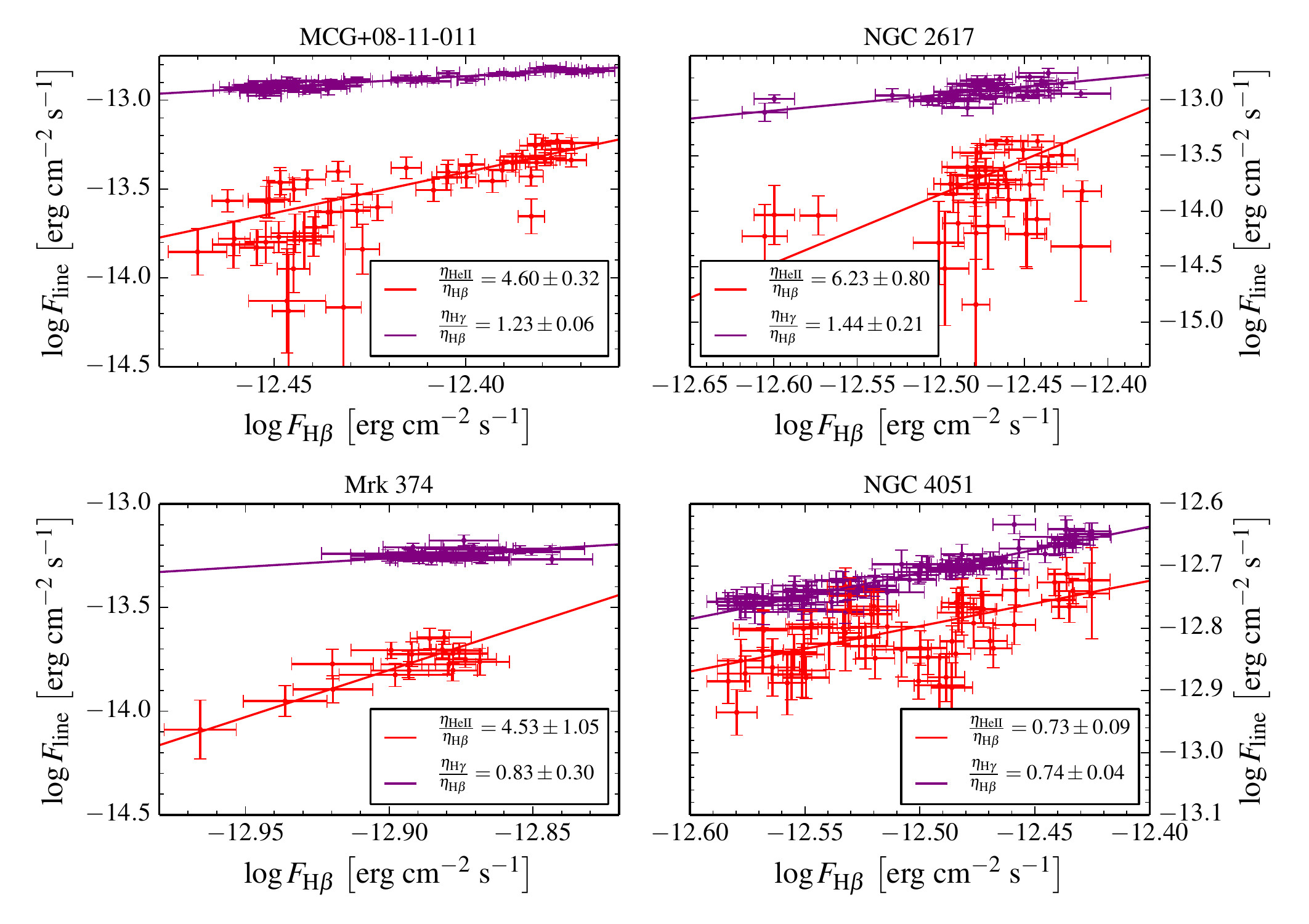}
\caption{Responsivities of optical recombination lines relative to
  H$\beta$.  The results for H$\gamma$ are shown in purple and those
  for \heii\ are shown in red.  Solid lines show weighted linear-least
  squares fits (accounting for uncertainties in both
  coordinates)---the slope of the fit gives the relative responsivity,
  which is listed in each panel.  See \S5.3 for more
  details.  \label{fig:responsivity}}
\end{figure*}
\subsection{Photoionization Physics}

Photoionization models make predictions about the structure of the BLR
that can be tested with RM of multiple recombination lines.  The
locally optimally emitting cloud model \citep{Baldwin1995} provides a
natural explanation for the general similarity of AGN spectra, and
predicts radial stratification of the BLR---high-ionization state
lines, such as He{\sc ii}\,$\lambda$1640/4686 and C{\sc iv}\,$\lambda
1549$, should be primarily emitted at smaller radii than
low-ionization state lines such as H$\beta$ and Mg{\sc ii}\,$\lambda
2798$. \citet[hereinafter KG04]{Korista2004} use this model to predict
that the responsivity of high-order Balmer lines should be greater
than that of low-order lines (in the sense that ${\rm H\gamma > H\beta
  > H\alpha}$).  KG04 also predict that high-ionization state lines
such as \heii\ should have greater responsivity than all of the Balmer
lines.  Radial stratification of the BLR in NGC\,5548 was first
observed by \citet{Clavel1991}, and has since been observed in several
other objects (\citealt{Peterson2004,Grier2013}).  In addition, the
expected trends of responsivity with ionization state/species have
been confirmed in 16 AGN by LAMP \citep{Bentz2010, Barth2015}.

We confirm these results for the four objects with multiple line light
curves presented here. The \heii\ lags in MCG+08-11-011 and NGC\,2617
are less than 2 days, while the H$\beta$ lags are 14.82 and 6.38 days,
respectively, clearly indicating radial stratification.  Furthermore,
the fractional variability of the light curves, as measured by $F_{\rm
  var}$ (Table\,\ref{tab:lc_prop}), is generally larger for H$\gamma$
than H$\beta$ (or comparable for NGC\,4051 and Mrk\,374), while
$F_{\rm var}$ for \heii\ is always much greater than for the Balmer
lines (although it is only slightly higher in NGC 4051).  This implies
that the relative line responsivities are \heii\ $\gg$ H$\gamma$ $>$
H$\beta$, in agreement with the photoionization models.  We also find
that the H$\gamma$ lags are slightly shorter than the H$\beta$ lags
within the same object (except for NGC\,4051).  KG04 show that shorter
lags are a natural consequence of the higher responsivity of H$\gamma$
compared to H$\beta$.

The formal definition of the responsivity of an emission line is
\begin{align}
  \eta_{\rm line} = \frac{\Delta \log F_{\rm line}}{\Delta \log \Phi}
\end{align}
where $F_{\rm line}$ is the line flux and $\Phi$ is the photoionizing
flux (KG04). The parameter $\eta_{\rm line}$ is therefore a measure of
how efficiently the BLR converts a {\it change} in the photoionizing
flux into a {\it change} in line emission.  The ionizing flux $\Phi$
cannot be observed directly because these photons are at far UV
wavelengths ($<912$\,\AA).  Therefore, we cannot measure $\eta_{\rm
  line}$ directly, but we can measure the relative responsivity
$\eta_{\rm line1}/\eta_{\rm line2} = \Delta \log F_{\rm line1}/\Delta
\log F_{\rm line2}$.

We present rough measurements of the relative responsivity of
H$\beta$, H$\gamma$, and \heii\ in Figure\,\ref{fig:responsivity}.
For each object, we first removed the lags of each line from the
corresponding light curve.  We then matched observed points to the
nearest day between the H$\beta$ light curves and H$\gamma$ or \heii\
light curves.  The ratio $\eta_{\rm line}/\eta_{\rm H\beta}$ then
corresponds to the slope of a linear least-squares fit to the data in
the $\log F_{\rm H\beta}$-$\log F_{\rm line}$ plane.

We find that $\eta_{\rm H\gamma}/\eta_{\rm H\beta}$ ranges from $0.74$
to $1.44$ and that $\eta_{\rm He{\sc II}}/\eta_{\rm H\beta}$ ranges
between $0.73$ and $6.23$.  NGC\,4051, with $\eta_{\rm He{\sc
    II}}/\eta_{\rm H\beta} \sim 0.73$, is an outlier, probably caused
by over-subtracting the continuum before integrating the line flux.
For comparison, KG04 calculate $\eta_{\rm line}$ for a fiducial model
of the BLR in NGC\,5548, which includes an empirically motivated but
{\it ad hoc} parameterization of the ionizing flux.  From their Table
1, $\eta_{\rm H\gamma}/\eta_{\rm H\beta}$ ranges between 1.03 and
1.07, depending on the flux state of the AGN, while $\eta_{\rm He{\sc
    II}}/\eta_{\rm H\beta}$ ranges from 1.26 to 1.61.  Thus, while our
fits for $\eta_{\rm H\gamma}/\eta_{\rm H\beta}$ are in reasonable
agreement with this fiducial model, the values of $\eta_{\rm He{\sc
    II}}/\eta_{\rm H\beta}$ are much larger than the model's
prediction.  The spread of $\eta_{\rm line}/\eta_{\rm H\beta}$ in our
fits is also fairly large, which may indicate a diversity of
photoionization conditions in the BLRs of different objects (perhaps
due to harder or softer ionizing fluxes than assumed for NGC\,5548).

Our estimates of the relative responsivities are sensitive to the
total flux of the line light curves.  For example, the sublinear
slopes for $\eta_{\rm H\gamma}/\eta_{\rm H\beta}$ in NGC\,4051 and
Mrk\,374 could be explained by missing variable line flux, perhaps in
the wings of the line during low-flux states, or excess constant flux
from the narrow emission lines or host-galaxy starlight.  On the other
hand, large values of $\eta_{\rm He{\sc II}}/\eta_{\rm H\beta}$ might
be explained by contamination by Fe{\sc ii} lines or misestimation of
the continuum.

\section{Summary and Future Prospects}
We have presented the initial analysis of data from an intensive RM
monitoring campaign carried out in the first half of 2014.  We
succeeded in measuring continuum-line lags for six targets, five of
which are presented here.  (For NGC\,5548, see \citealt{Pei2017}.) Our
main results are:
\begin{enumerate}[i.]
\item Four new SMBH masses, as well as a refined measurement for NGC
  4051.
\item In addition to measuring H$\beta$ lags for all five targets, we
  measure H$\gamma$ lags in four objects and \heii\ lags in two
  objects.
\item Using the \heii\ lags (or their upper limits), we show
  that the BLR is radially stratified.  Although the \heii\ virial
  products are somewhat smaller than those derived from H$\beta$,
 systematic effects such as blending in the line wings and
    the choice of continuum interpolation may account for these
    discrepancies. The BLRs are otherwise consistent with virialized
  dynamics with $V(r) \propto r^{-1/2}$.
\item We find that \heii\ is more responsive than the Balmer lines,
  and that H$\gamma$ is more responsive than H$\beta$, in agreement
  with predictions from photoionization modeling.
\end{enumerate}

Many modern RM experiments are focused on measuring velocity-resolved
reverberation signatures, in order to investigate the geometry and
dynamics of the BLR.  There are only six AGN with published
velocity-delay maps \citep{Ulrich1996, Bentz2010,Grier2013} and five
AGN with direct BLR dynamical models (\citealt{Pancoast2014b}; one
AGN, Arp 151, has both).  The data presented here are of exceptional
quality and very well-calibrated---based on the cadence and S/N of
these observations, we have an excellent prospect of recovering
velocity-delay maps and dynamical models in three objects
(MCG+08-11-011, NGC\,2617, and NGC\,4051).  This will expand the
sample of AGN with detailed BLR information by $\sim$30\%,
demonstrating the continuing importance of targeted and intensive
monitoring campaigns.

\acknowledgments

M.M.F. acknowledges financial support from a Presidential Fellowship
awarded by The Ohio State University Graduate School. NSF grant
AST-1008882 supported M.M.F., G.D.R., B.M.P., and R.W.P.,
M.C.B. gratefully acknowledges support through NSF CAREER grant
AST-1253702 to Georgia State University.  K.D.D. is supported by an
NSF AAPF fellowship awarded under NSF grant AST-1302093.  C.S.K. is
supported by NSF grant AST-1515876.  K.H. acknowledges support from
STFC grant ST/M001296/1.  This material is based in part upon work
supported by the National Science Foundation (NSF) Graduate Research
Fellowship Program under Grant No.\ DGE-0822215, awarded to C.B.H.
A.M.M. acknowledges the support of NSF grant AST-1211146.  M.E. thanks
the members of the Center for Relativistic Astrophysics at Georgia
Tech, where he was based during the observing campaign, for their warm
hospitality.  J.S.S. acknowledges CNPq, National Council for
Scientific and Technological Development, Brazil.  J.T. acknowledges
support from NSF grant AST-1411685.  Work by S.V.Jr. is supported by
the National Science Foundation Graduate Research Fellowship under
Grant No. DGE-1343012.  Work by W.Z. was supported by NSF grant
AST-1516842.  T.W.-S.H. is supported by the DOE Computational Science
Graduate Fellowship, grant number DE-FG02-97ER25308.  E.R.C. and
S.M.C. gratefully acknowledge the receipt of research grants from the
National Research Foundation (NRF) of South Africa. T.T. acknowledges
support by the National Science Foundation through grant AST-1412315
"Collaborative Research: New Frontiers in Reverberation Mapping," and
by the Packard Foundation through a Packard Research Fellowship.
D.J.S.  acknowledges support from NSF grants AST-1412504 and
AST-1517649.  A.J.B. and L.P. have been supported by NSF grant
AST-1412693.  B.J.S. is supported by NASA through Hubble Fellowship
grant HF-51348.001 awarded by the Space Telescope Science Institute
that is operated by the Association of Universities for Research in
Astronomy, Inc., for NASA, under contract NAS 5-26555.

This work makes use of observations with the NASA/ESA Hubble Space
Telescope. MCB acknowledges support through grant {\it HST} GO-13816
from the Space Telescope Science Institute, which is operated by the
Association of Universities for Research in Astronomy, Inc., under
NASA contract NAS5-26555.  This work is based on observations obtained
at the MDM Observatory, operated by Dartmouth College, Columbia
University, Ohio State University, Ohio University, and the University
of Michigan.  This paper is partly based on observations collected at
the Wise Observatory with the C18 telescope. The C18 telescope and
most of its equipment were acquired with a grant from the Israel Space
Agency (ISA) to operate a Near-Earth Asteroid Knowledge Center at Tel
Aviv University.  The Fountainwood Observatory would like to thank the
HHMI for its support of science research for undergraduate students at
Southwestern University.  This research has made use of NASA's
Astrophysics Data System, as well as the NASA/IPAC Extragalactic
Database (NED) which is operated by the Jet Propulsion Laboratory,
California Institute of Technology, under contract with the National
Aeronautics and Space Administration.

\facility{McGraw-Hill, HST, Wise Observatory, Fountainwood Observatory
  BYU:0.9m, CrAO:0.7m, WIRO, LCOGT, SSO:1m}

\software{ Astropy \citep{astropy},
  IRAF \citep{iraf},
  Matplotlib \citep{matplotlib},
  Numpy \citep{numpy},
  Scipy \citep{scipy} }

\bibliography{../../refs}

\end{document}